\documentclass[a4paper,12pt,superscriptaddress, floatfix,showpacs,article]{revtex4}

\usepackage{graphicx}
\usepackage{subfigure}
\usepackage{bm}% bold math
\usepackage{color}
 \expandafter\let\csname equation*\endcsname\relax
  \expandafter\let\csname endequation*\endcsname\relax
\usepackage{amsmath,amssymb,amsfonts}
\usepackage{verbatim}% Long comments

\DeclareMathOperator{\diag}{diag}

\begin{document}

\title[Role of local geometry]{Role of local geometry in spin and orbital structure of transition metal compounds}

\author{D.~I.~Khomskii} \affiliation{$II.$ Physikalisches
Institut, Universit\"at zu K\"oln, Z\"ulpicher Str. 77, 50937
K\"oln, Germany}

\author{K.~I.~Kugel} \affiliation{Institute for Theoretical and
Applied Electrodynamics, Russian Academy of Sciences, Izhorskaya
Str. 13, Moscow, 125412 Russia}

\author{A.~O.~Sboychakov} \affiliation{Institute for Theoretical and
Applied Electrodynamics, Russian Academy of Sciences, Izhorskaya
Str. 13, Moscow, 125412 Russia}

\author{S.~V.~Streltsov} \affiliation{Institute of Metal Physics,
Ural Branch, Russian Academy of Sciences, S. Kovalevskaya Str. 18,
Ekaterinburg, 620990 Russia} \affiliation{Ural Federal University,
Mira Str.  19, Ekaterinburg, 620002 Russia}

\date{\today}

\begin{abstract}
We analyze the role of local geometry in the spin and orbital interaction in transition metal compounds with orbital degeneracy. We stress that the tendency observed for the most studied case (transition metals in O$_6$ octahedra with one common oxygen -- common corner of neighboring octahedra and with $\sim 180^{\circ}$ metal--oxygen--metal bonds), that ferro-orbital ordering renders antiferro-spin coupling, and, {\it vice versa}, antiferro-orbitals give ferro-spin ordering, is not valid in general case, in particular for octahedra with common edge and with $\sim 90^{\circ}$ M--O--M bonds. Special attention is paid to the ``third case'', neighboring octahedra with common face (three common oxygens) -- the case practically not considered until now, although there are many real systems with this geometry. Interestingly enough, the spin--orbital exchange in this case turns out to be to be simpler and more symmetric than in the first two cases. We also consider, which form the effective exchange takes for different geometries in case of strong spin--orbit coupling.

\end{abstract}

\pacs{75.25.Dk, %Orbital, charge, and other orders, including coupling of these orders
75.30.Et, %Exchange and superexchange interactions
75.47.Lx, %Magnetic oxides
71.27.+a, %Strongly correlated electron systems
71.70.Ej, %Spin-orbit coupling, Jahn-Teller effect
75.10.Dg  %Crystal-field theory and spin Hamiltonians
}

\maketitle

\section{Introduction \label{intro}}

The study of correlated systems  with orbital ordering (OO) is currently a very active field of research in solid state physics. Orbital ordering is not only accompanied (or caused) by structural transitions, but it also largely determines magnetic properties of many materials, e.g., transition metal (TM) oxides: according to the Goodenough--Kanamori--Anderson  rules~\cite{Gooden_book} the orbital occupation largely determines the magnitude and even sign of exchange interaction. By modifying orbital occupation one can control magnetic properties of a system~\cite{ChakhalScience2007}. Besides more traditional electron-lattice (Jahn--Teller) mechanism~\cite{KaplVekh_book} of OO, also a purely electronic (exchange) mechanism can lead to both orbital and magnetic
ordering~\cite{KugKhoUFN82}, which appear to be coupled.

The coupled spin and orbital ordering depends not only on electronic structure of constituent ions, but also on the local geometry of the system. The most often treated case is the system with a transition metal ion (M) surrounded by the ligand (e.g., oxygen, O) octahedra, with these neighboring MO$_6$ octahedra having  one common oxygen (common corner) with the M--O--M angle of  about $180^{\circ}$ (it may also be smaller than $180^{\circ}$, but small deviations of this angle from  $180^{\circ}$ do not play an important role, see, e.g., book~\cite{Khomskii_book2014}). This situation is met, for example, in such important systems as perovskites, e.g., CMR manganites (LaSr)MnO$_3$ or in high-T$_c$ cuprates like (LaSr)$_2$CuO$_4$. For this case, one knows a lot: what is the form of electron--lattice (Jahn--Teller) interaction leading to the orbital ordering and how the spin exchange looks like.  The general conclusion reached in the study of these systems is that ferro-orbital ordering usually leads to the antiferromagnetic spin ordering, and {\it vice versa} antiferro-orbital ordering gives rise to the ferromagnetic spin exchange. This ``rule'' became a kind of ``folklore'' and is used by many theoreticians and experimentalists to explain or predict the type of coupled spin and orbital ordering in various systems with different crystal structure. However, one has to realize that this ``rule'' was derived for this particular geometry, and it does not have to be fulfilled in other cases. Thus, for example, it is not the case for neighboring MO$_6$ octahedra having two common oxygens (common edge) with the M--O--M angle of about $90^{\circ}$.  There are a lot of interesting and important materials with this local geometry, e.g., the ``battery material'' LiCoO$_2$, many frustrated systems, multiferroics, etc. (one should note right away that here the deviations from the ``pure'' $90^{\circ}$ M--O--M angle may have more drastic consequences). As we will discuss below (see also \cite{Khomskii_book2014,StrelKhomsPRB2014orb-sel}), thus general ``rule'' is strongly violated in this case.

There exists also much less studied ``third case'' of neighbors with three common oxygens, the systems with MO$_6$ octahedra sharing common face. To understand the systematics of coupled spin and orbital ordering in these different cases is an interesting and practically important problem, which will be discussed in this paper.

In addition to the local M--O geometry, which is crucial for the superexchange originating from the virtual hopping of $d$ electrons via ligands, e.g. oxygens, in the second and the third cases (common edge and common face), in contrast to the simpler case of common corner, electron hopping leading to superexchange can occur not only via ligands, but there may exist also significant direct overlap of certain $d$ orbitals of neighboring TM ions. The resulting  direct exchange should be also taken into account in certain cases; it can lead to very nontrivial effects ~\cite{StrelKhomsPRB2014orb-sel}.

One more factor, important especially for heavy ($4d$, $5d$) transition metals and attracting much attention nowadays, is the role of the relativistic spin--orbit coupling (SOC), which for these elements could be very significant and which could sometimes dominate the properties of corresponding systems. In this paper, we consider these effects for different geometries.

The plan of the paper is the following. In Sections 2 and 3, we shortly summarize the known properties of spin and orbital exchange for the geometries with common corner (M--O--M angle of about $180^{\circ}$) and common edge ($90^{\circ}$ M--O--M bonds). In particular, we want to stress here several important features, which distinguish these two cases. We also compare the situation in these two cases for systems with strong spin--orbit coupling. Then, in Sections 4 and 5, we discuss the much less studied case of systems with common face of neighboring MO$_6$ octahedra, using partially the results of our recent paper~\cite{MyPRB2015_SU4}, generalizing those for the case a finite noncubic crystal field and also for that of significant spin--orbit coupling.

\section{Systems with the common-corner geometry}

\begin{figure}[t]
\centering
\includegraphics[width=0.6\columnwidth]{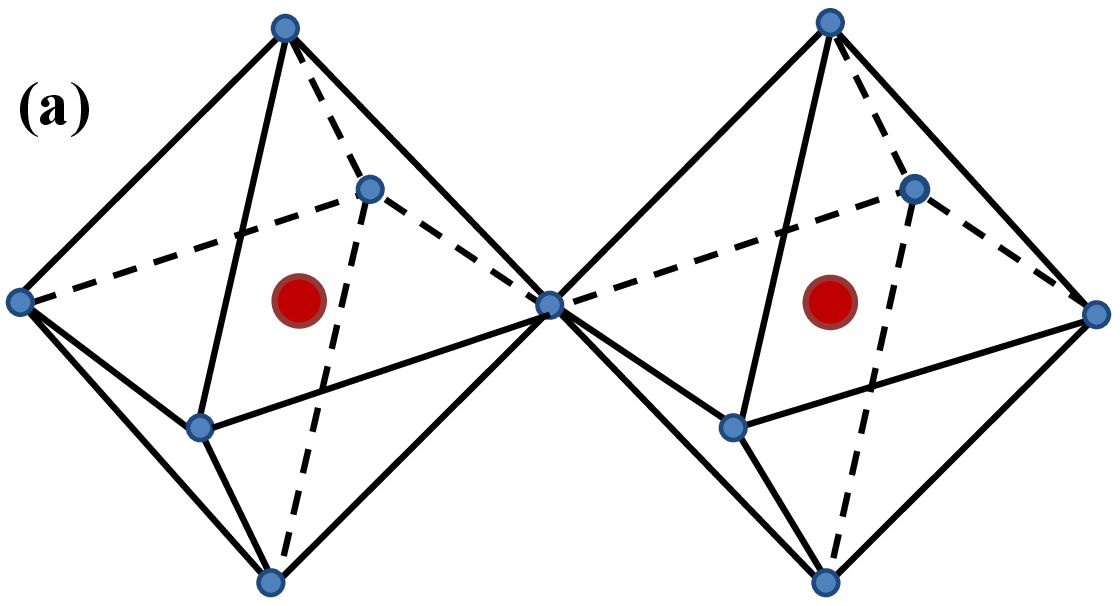}
\includegraphics[width=0.35\columnwidth]{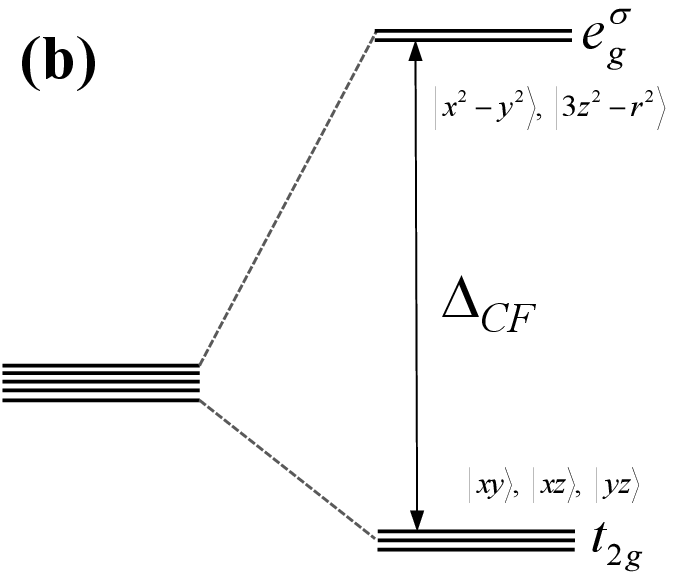}
\caption{(Color online) (a) Corner-sharing octahedra. Large (red) and small (blue) circles denote metal and ligand ions, respectively.  (b) Crystal field splitting of $d$ orbitals of the metal ion. } \label{corner_sharing}
\end{figure}

The geometry of the systems with MO$_6$ octahedra sharing a common corner is illustrated in Fig.~\ref{corner_sharing}(a). Here, the transition metals are rather far away from each other, with a ligand (oxygen) ions in between. Thus, in such situation, all electron hoppings occur via oxygen. We also remind in Fig.~\ref{corner_sharing}(b) what is the crystal field (CF) splitting in the case of ideal octahedra. For the  tetragonal reference frame with axes directed from M to O ions,  the $e_g$  orbitals are $|x^2-y^2\rangle$ and $|3z^2-r^2\rangle$, while $t_{2g}$ orbitals are $|xy\rangle$, $|yz\rangle$, and $|zx\rangle$.

For the simple case of one electron or one hole in a doubly degenerate $e_g$ orbital (Cu$^{2+}$ or low-spin Ni$^{3+}$), we can describe the state of an ion by the spin $S=1/2$ and by the orbital occupation, which can be also mapped into pseudospin-1/2 situation, with pseudospin projection $\tau^z=1/2$ corresponding to orbital $1$, say $|x^2-y^2\rangle$, and $\tau^z = -1/2$ -- to orbital 2, $|3z^2-r^2\rangle$. One can also make arbitrary linear superpositions of these states, of the type $\alpha |3z^2-r^2 \rangle + \beta | x^2-y^2 \rangle$, $|\alpha|^2 + |\beta|^2 =1 $, where coefficients $\alpha$ and $\beta$ can in principle be complex \cite{Brink2001}. For one electron per site and for the strongly interacting case (with the Hubbard on-site electron repulsion $U$ far exceeding the electron hopping integral $t$) the usual treatment in perturbation theory in $t/U$ leads to the following schematic form of the exchange Hamiltonian
\begin{equation} \label{KK_model}
 H =J_1\sum_{\langle ij\rangle}\mathbf{S}_i\mathbf{S}_j +\sum_{\langle ij\rangle}J_{2ij}^{\alpha\beta}{\tau}^{\alpha}_i{\tau}^{\beta}_j +\sum_{\langle ij\rangle} J^{\alpha\beta}_{3ij}(\mathbf{S}_i\mathbf{S}_j)({\tau}^{\alpha}_i{\tau}^{\beta}_j)\,,
\end{equation}
where the summation is taken over the nearest-neighbor sites, $\mathbf{S}_i$  is the spin of site $i$, and the pseudospin operators $\bm{\tau}_i$ describe orbital state for the case of double orbital degeneracy (say, for $e_g$ levels). The spin part of this exchange is of a Heisenberg type ($\mathbf{S}_i\mathbf{S}_j$), but the orbital part may be more complicated, containing anisotropic terms like $\tau_i^z \tau_j^z, \tau_i^x \tau_j^x, \tau_i^z \tau_j^x$, etc., which, in addition, depend on the relative orientation of the sites $i$ and $j$. Only in a more symmetric model (not actually realized for $e_g$ states) with the effective $d$-$d$ hopping such that $t_{11}=t_{22}=t,\, t_{12}=0$ the effective Hamiltonian takes the simpler symmetric form
\begin{equation} \label{symm_model}
H=J\sum_{\langle ij\rangle}\left(\frac{1}{2} + 2\mathbf{S}_i\mathbf{S}_j\right)\left(\frac{1}{2} + 2{\bm \tau}_i{\bm \tau}_j\right)\,,
\end{equation}
which has a rather high, not only SU(2)$\times$SU(2), but even SU(4) symmetry~\cite{AffleckSU4,YamashitaSU4,FrischmuthSU4}. Similarly, one can obtain the effective spin--orbital model for triply-degenerate $t_{2g}$ electrons, which, instead of pseudospin-$1/2$ orbital operators ${\bm\tau}_i$ would contain effective orbital $l=1$ operators, $\mathbf{l}_i$, describing three $t_{2g}$ states~\cite{KuKhFTT1975}. In general case, these operators would enter not only as $\mathbf{l}_i\mathbf{l}_j$, but with invariants of the type $l_i^zl_j^z-2/3$ or $l_i^xl_j^y+ l_j^yl_j^x$, etc.

As mentioned in the Introduction, for the doubly degenerate case in simple lattices such as that of perovskites AMO$_3$ with MO$_6$ octahedra having common corner [one common oxygen, with the M--O--M angle $\approx180^{\circ}$, see Fig.~\ref{corner_sharing}(a)], the typical situation is that the ferro-orbital ordering gives rise to the antiferromagnetic spin exchange, whereas antiferro-orbital ordering is rather favorable for spin ferromagnetism. However, one has to realize that this conclusion was reached for particular cases, for systems with particular geometry, with $180^{\circ}$ M--O--M superexchange, and the situation is very different for other cases, as will be demonstrated below.

We conclude this section by presenting the results for this geometry in the case of  the strong spin--orbit coupling. The SOC is quenched for $e_g$ electrons, see, e.g.~\cite{Gooden_book,Khomskii_book2014}, but the interesting and nontrivial results can occur for partially filled $t_{2g}$ levels. However, just this is the typical situation for 4$d$ and 5$d$ compounds for which SOC is strong: due to large CF splitting $\Delta_{CF} = 10Dq$ and smaller Hund's energy (Hund's rule coupling $J_H$ is $\sim 0.8-0.9$\,eV for 3$d$ elements, $\sim 0.6-0.7$\,eV for 4$d$, and $\sim 0.5$\,eV for 5$d$) these ions are usually in the low-spin state, i.e. their electrons first fill $t_{2g}$ levels, and only for $n_d>6$ electrons start to occupy $e_g$ levels. For these cases,  sometimes one can project electronic states to the ground-state multiplets calculated including SOC. Here, one has to discriminate between the cases with different electron occupation (less than half-filled $t_{2g}$ shell, $n_{t_{2g}}<3$, and more than half-filled shell, $n_{t_2g} >3$). The second (or the third) Hund's rule for partially-filled $t_{2g}$ shells tells us that the ground state multiplet for $n_{t_{2g}}<3$ corresponds to the maximum possible total momentum $j$, and for the case of more than half-filled $t_{2g}$ shell -- to minimum possible $j$. In effect, e.g. for dominating SOC, the ions with the $d^5$ configuration (Ir$^{4+}$, Os$^{3+}$) would have as a ground state a Kramers doublet $j=1/2$ (for $t_{2g}$ triplet $l_{eff}=1$, $S=1/2$, which gives $j=1/2$ doublet as a ground state), which can be described by the effective spin $\sigma = 1/2$ (usual Pauli matrices)~\cite{Khomskii_book2014,AbragamBleaney1970}. The superexchange then can be projected onto this subspace and written through pseudospin $j=1/2$ operators. The form of this exchange for strongly localized electrons ($U>{J_H} \gg t$) has been obtained for perovskites with 180$^{\circ}$ exchange, e.g. Sr$_2$IrO$_4$, and for 90$^{\circ}$ exchange, e.g. for Na$_2$IrO$_3$, in Ref.~\cite{JackKhal_PRL2009}. The resulting exchange interaction in terms of $j=1/2$ looks very different: it is predominantly Heisenberg-like for $180^{\circ}$ bonds, but is strongly anisotropic (Ising-like) for each Ir--O--Ir pair in case of IrO$_6$ octahedra with common edge oxygens. Indeed, for the case of corner-sharing octahedra, the dominant exchange has a simple Heisenberg form~\cite{JackKhal_PRL2009}
\begin{equation} \label{JackKhal}
 H =J\sum_{\langle ij\rangle}{\bm \sigma}_i{\bm \sigma}_j\,,
 \end{equation}
where ${\bm \sigma}$ are the Pauli matrices describing  effective spin 1/2 for the $j=1/2$ Kramers doublets. We see that the strong spin--orbit coupling can effectively remove orbital degeneracy, and instead of the complicated spin--orbital Hamiltonian of the type of  Eqs.~\eqref{KK_model} or~\eqref{symm_model} (or even more complicated form for the triple $t_{2g}$ degeneracy, see above and Ref.~\cite{KuKhFTT1975}), one obtains the simple Heisenberg interaction~\eqref{JackKhal}. However, whether in real cases we indeed meet the situation, in which the SOC really dominates, is a special question, which should be addressed for each specific system.

\section{Spin-orbital exchange for the octahedra with common edge}

\begin{figure}[t]
\centering
\includegraphics[width=0.5\columnwidth]{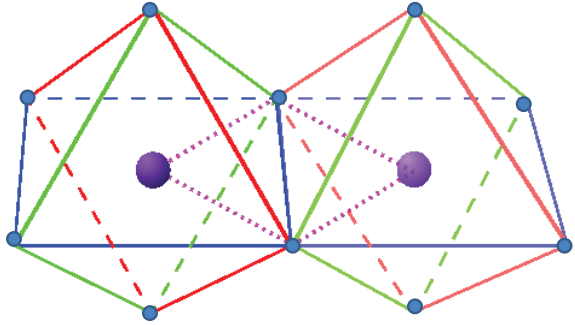}
\caption{(Color online) Edge-sharing octahedra. Large and small circles denote metal and ligand ions, respectively.} \label{edge_sharing}
\end{figure}

\begin{figure}[t]
\centering
\includegraphics[width=0.30\columnwidth]{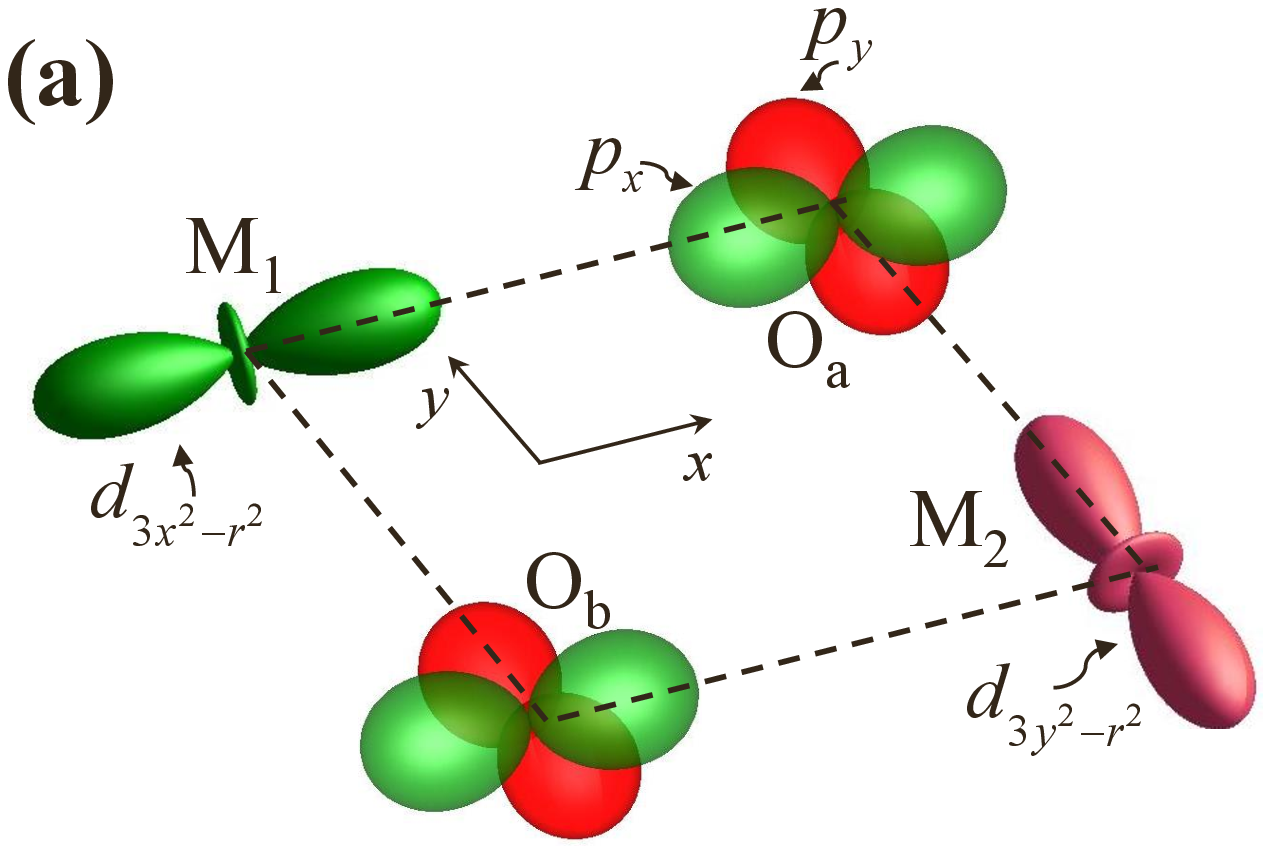}
\includegraphics[width=0.30\columnwidth]{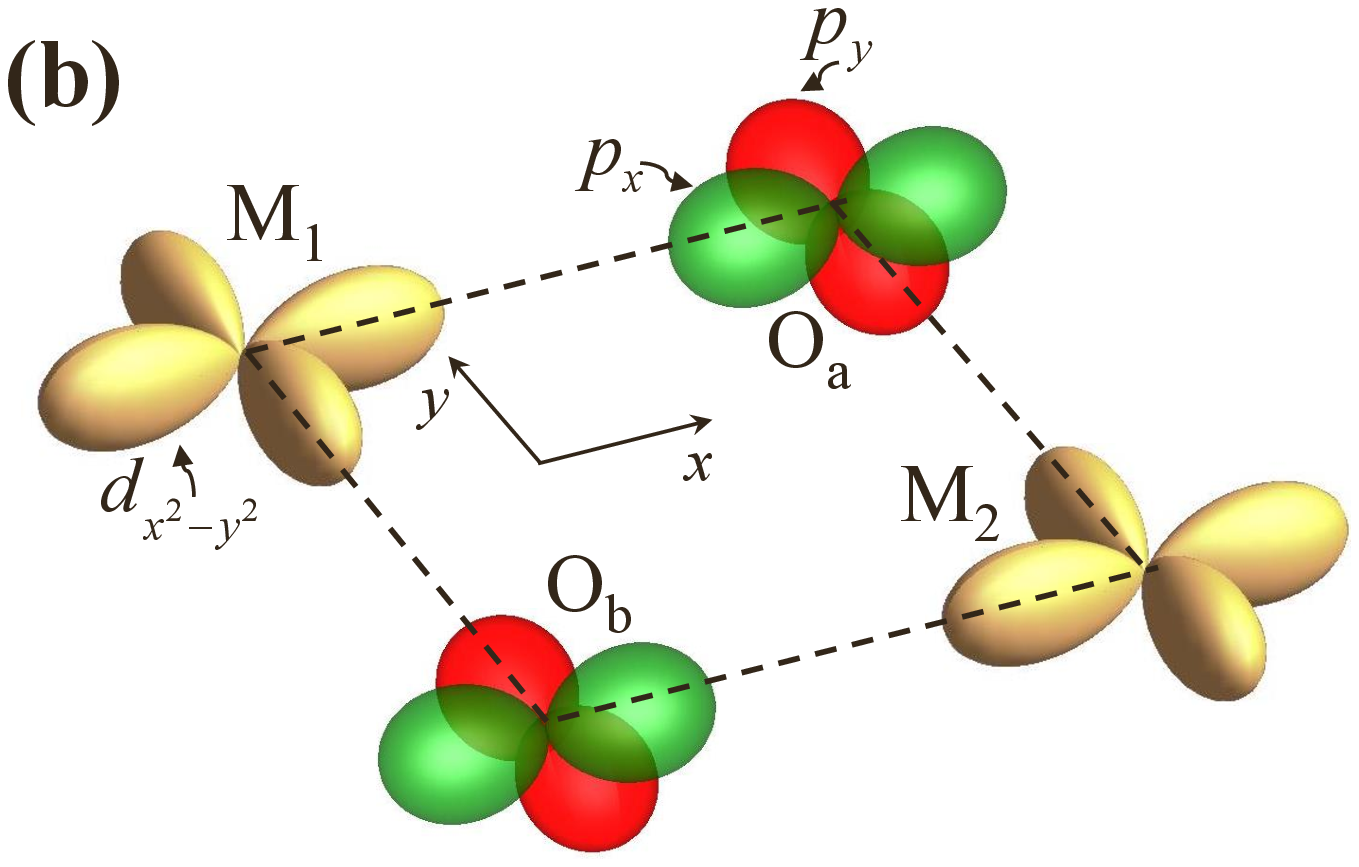}
\includegraphics[width=0.30\columnwidth]{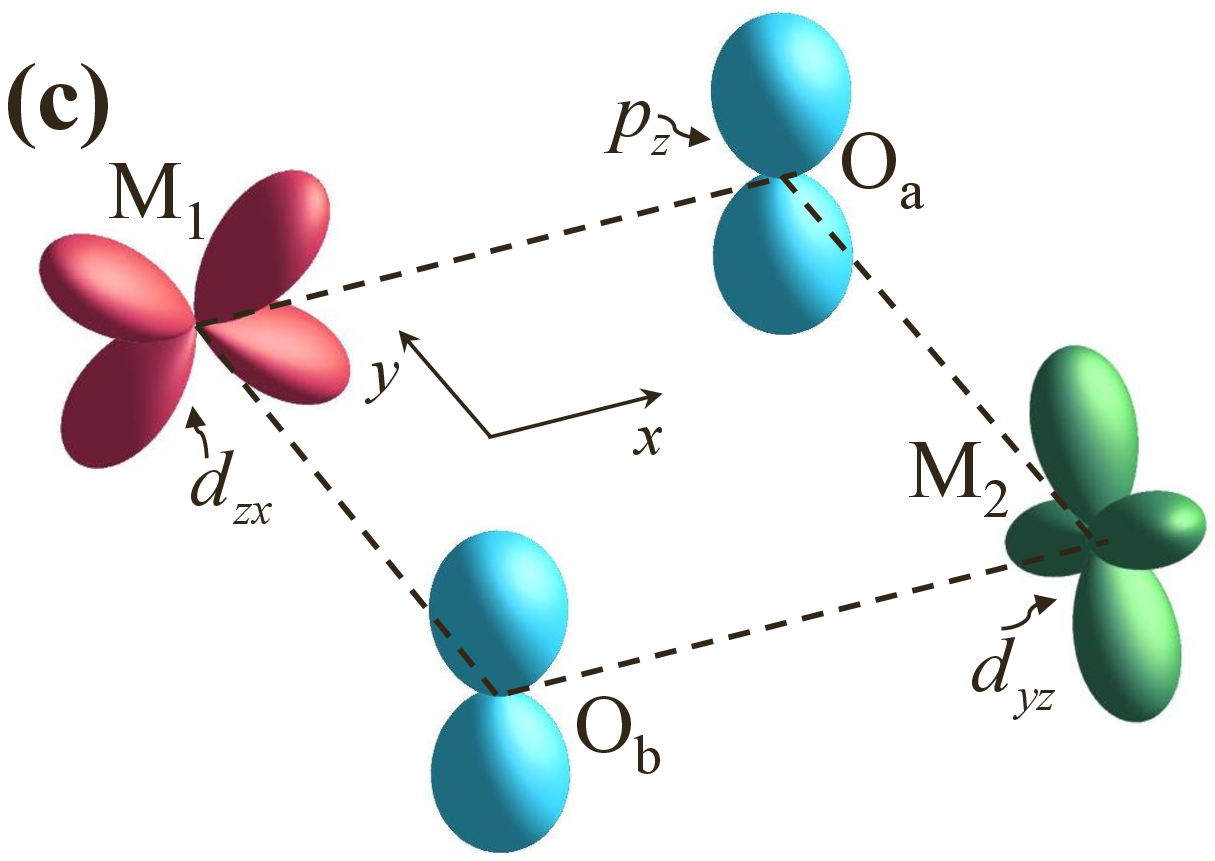}
\caption{(Color online) (a) ``Active''  orbitals contributing to the M$_1$-O$_a$-M$_2$ superexchange for the 90$^{\circ}$ metal-oxygen-metal bonds typical for the common edge geometry. (b) Ferro-orbital ordering of $|x^2-y^2\rangle$ orbitals in the $xy$ plane in the case of edge-sharing octahedra. (c) Antiferro-orbital ordering of $|yz\rangle$ and $|zx\rangle$ orbitals in the $xy$ plane in the case of edge-sharing octahedra.}\label{anti-ferro-orbit}
\end{figure}

Another typical situation in TM compounds is that with neighboring TM ions having two common oxygens; for systems with MO$_6$ octahedra this is the case of common edge, with  $\approx90^{\circ}$  M--O--M bonds, see Fig.~\ref{edge_sharing} (note that here instead of oxygens there may be other ligands: halogens such as F, Cl, or S, Se, Te, etc.). This situation is typical, for example, for B sites of spinels, or in layered materials with CdI$_2$ or with delafossite structures, etc. In this case, the situation with spin--orbital exchange is quite different from that of octahedra with common corner, see e.g. \cite{MostKhomsPRL2002} and discussion below. Thus, for example for $e_g$ electrons, the exchange interaction is ferromagnetic both for ferro- and for antiferro-orbital ordering. The effective superexchange interaction in the case of doubly degenerate $e_g$  orbitals would have schematically the form~\cite{MostKhomsPRL2002} (in the symmetric case)
\begin{equation}\label{MostKhoms_model}
H_{12} =-\tilde{J}\left(\frac 34+{\bm S}_1{\bm S}_2\right)\left[\left(\frac 12 + T^z_{x,1}\right) \left(\frac 12+ T^z_{y,2}\right) + \left(\frac 12 + T^z_{y,1}\right) \left(\frac 12 + T^z_{x,2}\right)\right]\, ,                                                                                                                        \end{equation}
where the first multiplier is the projection operator to a ferromagnetic (spin-triplet) state of a dimer M$_1$M$_2$, and the first term in square brackets is the projection to the orbitals $|3x^2-r^2\rangle$ and $|3y^2-r^2\rangle$ at sites 1 and 2, respectively, see Fig.~\ref{anti-ferro-orbit}(a). Here, $T^z_x$ and $T^z_y$ are the operators corresponding to $|3x^2-r^2\rangle$ and $|3y^2-r^2\rangle$  orbitals
\begin{eqnarray}
T^z_x = \frac{1}{2}\tau^z  - \frac{\sqrt3}{2}\tau^x\, ,\nonumber\\
T^z_y = \frac{1}{2}\tau^z  + \frac{\sqrt3}{2}\tau^x. \label{T-tau}
\end{eqnarray}
Only these orbitals overlap with  $|p_x\rangle$ and $|p_y\rangle$ orbitals of oxygens O$_\text{a}$ (and ``reversed'' orbitals $|3y^2-r^2\rangle$ on the site M$_1$ and $|3x^2-r^2\rangle$ on site M$_2$ with $p$ orbitals of O$_\text{b}$) and contribute to exchange; orthogonal orbitals do not overlap with $p$ orbitals and do not contribute to the exchange  (e.g. $|y^2-z^2\rangle$ orbital at a site M$_1$  is orthogonal to all $p$ orbitals of oxygen O$_\text{a}$ in Fig.~\ref{anti-ferro-orbit}(a)). In effect, the spin exchange turns out to be ferromagnetic for {\it any} ordering of $e_g$ orbitals.

This fact is also illustrated in Fig.~\ref{anti-ferro-orbit}(b) for $|x^2-y^2\rangle$ {\it ferro-orbital ordering}. This orbital from the site M$_1$ overlaps with the $|p_x\rangle$ orbital of oxygen O$_\text{a}$ (and with $|p_y\rangle$ orbital of oxygen O$_\text{b}$), whereas the same $d$ orbital of the TM site M$_2$ overlaps with the orthogonal $|p_y\rangle$ orbital of this oxygen (and with $|p_x\rangle$ orbital of oxygen O$_\text{b}$). In effect, we have ferromagnetic spin exchange of these two ions, stabilized by the Hund's rule exchange at an oxygen, $J_{H,p}$, when we virtually move two electrons from this oxygen to the TM sites M$_1$ and M$_2$, so that the resulting ferromagnetic exchange constant would be
\begin{equation}
\tilde{J}\sim\frac{t_{pd}^4}{\Delta_{CT}^2(\Delta_{CT}+U_p/2)}
\frac{J_{H,p}}{(\Delta_{CT}+U_p/2)}\,,
\end{equation}
where $t_{pd}$ is the metal--oxygen $p$--$d$ hopping amplitude, $\Delta_{CT}$ is the charge-transfer energy needed to move an electron from oxygen to metal ion, $d^np^6\rightarrow d^{n+1}p^5$, and $U_p$ is the on-site Coulomb repulsion energy of $p$ electrons at an oxygen site. Such a situation occurs  for instance in Mn$_3$O$_4$ spinel, with Mn$^{3+}$ ion at octahedral sites: the B--B exchange here is ferromagnetic, despite the ferro-OO  (in this case, the occupied $e_g$ orbitals are $|3z^2-r^2\rangle$) with the corresponding strong tetragonal elongation with $c/a\sim1.15$). One has to mention that this result is valid for Mott-Hubbard insulators, and it should
be modified for the charge-transfer insulators \cite{ZSA_PRL1985chtrgap}, see e.g. \cite{JackKhal_PRL2009} and \cite{Chen2008}.

One should also note that, in contrast to the $e_g$ case, for $t_{2g}$ electrons in the case of $90^{\circ}$ M--O--M bonds, we can have both ferro- and antiferromagnetic exchange, depending on the orbital occupation and on the exchange pass. Thus, for antiferro-orbital ordering with orbitals shown in Fig.~\ref{anti-ferro-orbit}(c) (occupied orbitals are $|zx\rangle$ and $|yz\rangle$ on the M$_2$O$_2$ plaquette in the $xy$ plane), the overlap with the $|p_z\rangle$ orbital of the oxygen O$_\text{a}$ would give strong antiferromagnetic spin exchange with
\begin{equation}
\tilde{J}\sim\frac{t_{pd}^4}{\Delta_{CT}^2}\left[\frac 1 U + \frac 1 {\Delta_{CT} + U_p/2}\right]\,.
\end{equation}
However, for different orbital occupation the exchange could again be ferromagnetic (for details, see, e.g., Ref.~\cite{Khomskii_book2014,StrKhomPRB2008pyroxenes}).

Let us notice also that for the case of $90^{\circ}$ M--O--M bonds the direct $d$-$d$ hopping amplitude, $t_{dd}$, for example for $|xy\rangle$ orbitals lying in the $xy$ plane of Fig.~\ref{anti-ferro-orbit}(c) and pointing directly toward one another (along the diagonal on M$_2$O$_2$ plaquette), may be quite significant and give contribution to the antiferromagnetic exchange $J$ of about $t_{dd}^2/U$.

The situation with common edge is also much richer and more complicated in comparison to that for the common corner in the case of strong SOC. Again, as we discussed above, the nontrivial effects appear for partially filled $t_{2g}$ shells, and one has to consider separately the cases with less than half-filled $t_{2g}$ shell, $n_{t_{2g}} < 3$, and more than half-filled shell, $n_{t_{2g}} > 3$. As mentioned above, in the first case, we have an inverted multiplet order, with the multiplet with maximum $j$ (and maximum degeneracy) lying lower in energy. This case is more difficult to consider technically, but at present it also attracts less attention, even though one might expect some interesting properties in this case as well. However, the main attention is attracted nowadays to the second case, e.g., to the systems with Ir$^{4+}$, Ru$^{3+}$ ($d^5$) and Ir$^{5+}$, Ru$^{4+}$ ($d^4$), where the ground state of an isolated ion is the state with minimum $j=1/2$ for $d^5$ and nonmagnetic $j=0$ state for $d^4$ configuration. How good is the limit of isolated ions for concentrated systems, in which the intersite electron hopping may become comparable or even exceed the SOC, is a very important question. However, if the atomic limit can be used as a valid starting point and the electron hopping can be treated as a small perturbation, one would obtain for the common edge geometry a very nontrivial result for $d^5$ ions (like Ir$^{4+}$): the resulting exchange projected to $j=1/2$ states has not Heisenberg, but rather an Ising form~\cite{JackKhal_PRL2009} of the type of  $\sigma^z\sigma^z$, where the $z$ axis is the direction perpendicular to the plane of M$_2$O$_2$ (e.g., Ir$_2$O$_2$) plaquette. For systems with honeycomb lattices like Na$_2$IrO$_3$~\cite{Singh_Na2IrO3_PRB2010}  or RuCl$_3$~\cite{PlumbRuCl3_PRB2014} the resulting exchange contains different effective spin combinations ($\sigma^x\sigma^x$, $\sigma^y\sigma^y$, and $\sigma^z\sigma^z$) for different bonds, so that in effect these compounds may be an example of what is known as system with the Kitaev interaction~\cite{Kitaev} -- a particular case of the so called compass model~\cite{KugKhoUFN82} (see also~\cite{Nussinov}). From our perspective, we again see that strong spin--orbit coupling acts against the usual (Jahn--Teller) orbital ordering, reducing initial degeneracy of the system in different way than the usual orbital ordering does.

\section{Spin--orbital exchange for the octahedra with common face}

\begin{figure}[t]
\centering
\includegraphics[width=0.3\columnwidth]{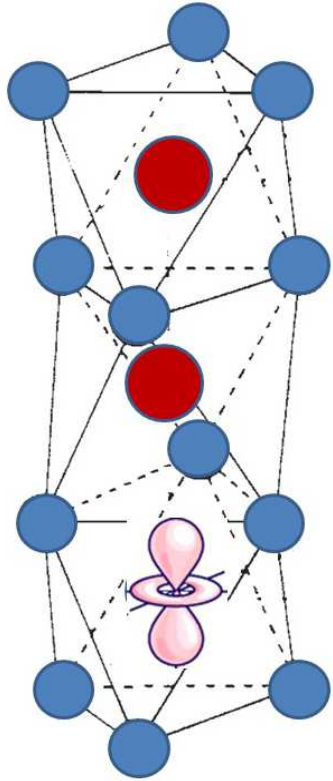}
\caption{(Color online) A chain of face-sharing octahedra. Large and small circles denote metal and ligand ions, respectively. The $a_{1g}$ orbital, which has strong direct overlap with similar orbital on the neighboring site, is shown for one transition metal ion.} \label{face_sharing}
\end{figure}

\begin{figure}[t]
\centering
\includegraphics[width=4.9 cm, height=7cm]{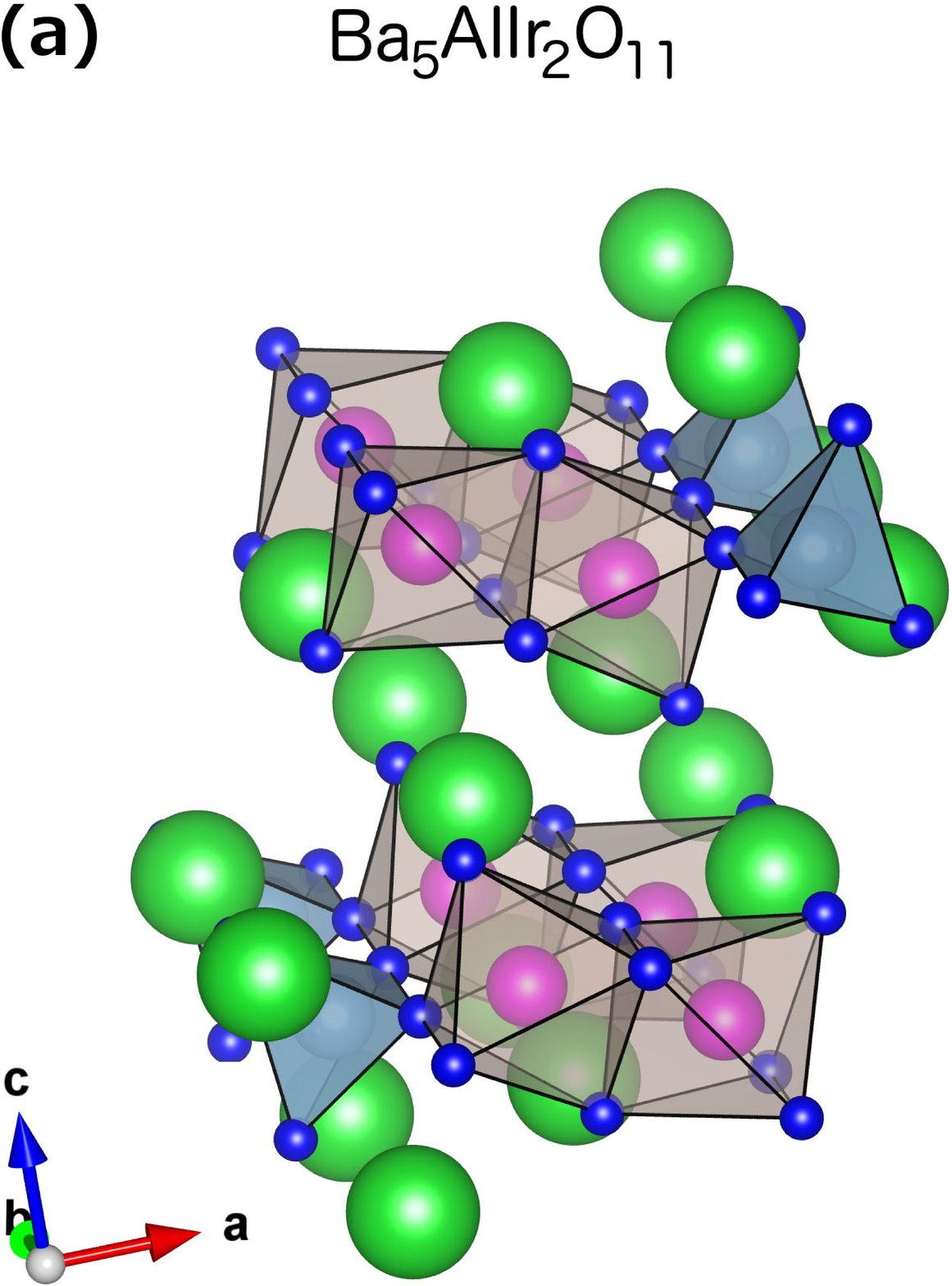}
\includegraphics[width=4.9 cm, height=7cm]{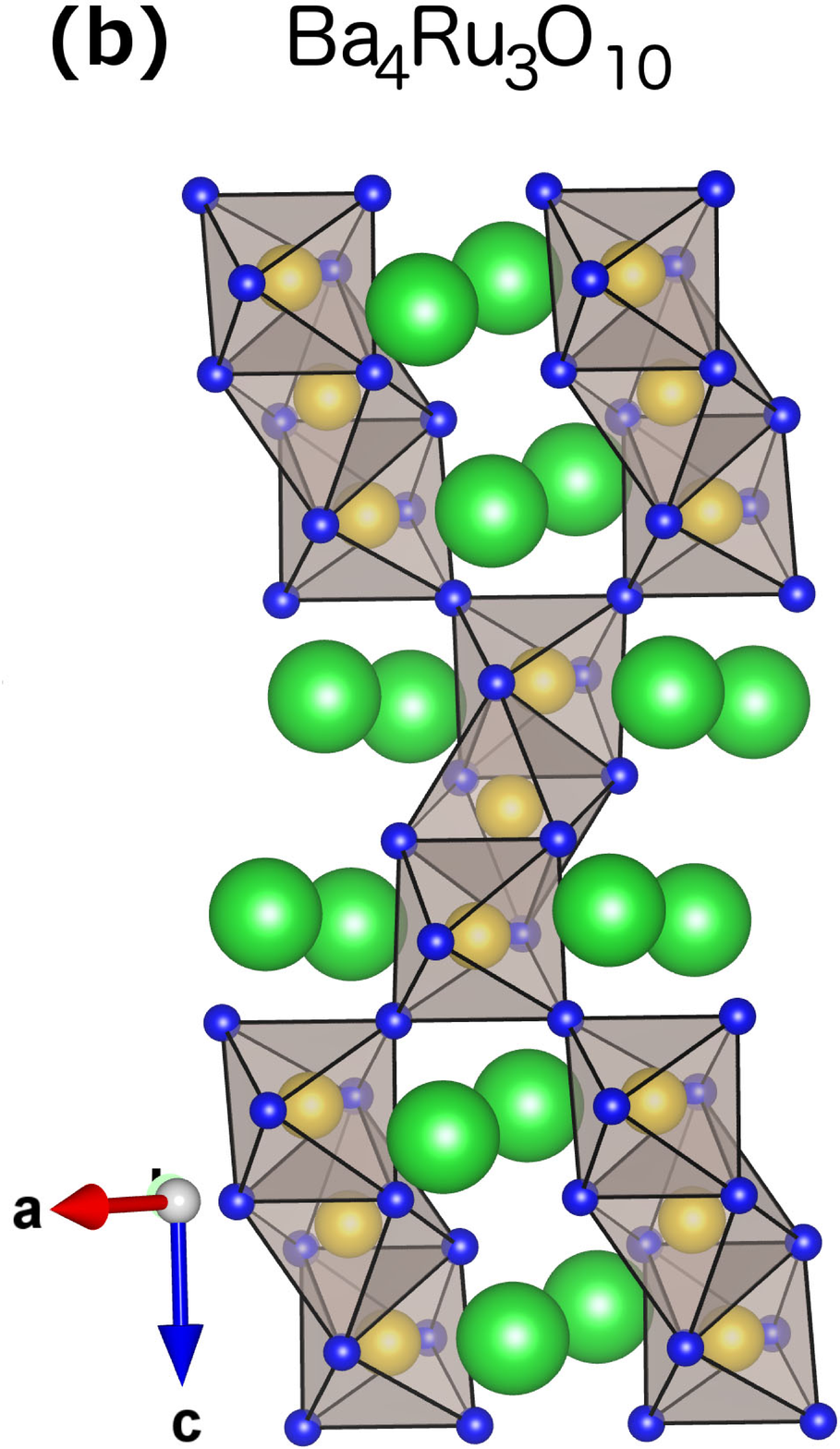}
\includegraphics[width=4.9 cm, height=7cm]{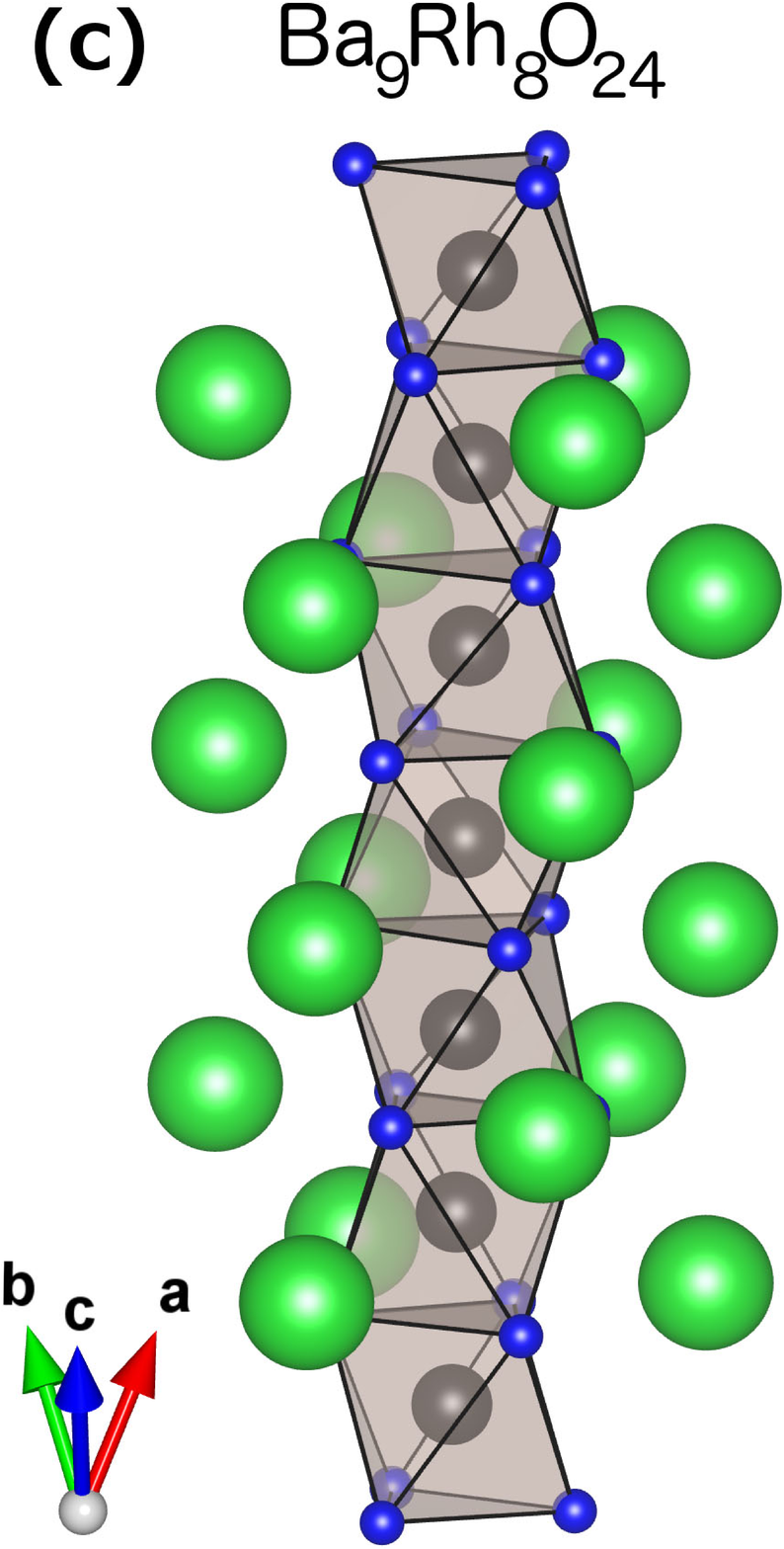}\caption{(Color online) Different arrays of face-sharing octahedra in various transition metal compounds.} \label{face-sharing_arrays}
\end{figure}

Both situations, with MO$_6$ octahedra with common corners (one common ligand, $\sim 180^{\circ}$ M--O--M bonds) and with common edge (two common oxygens, $\sim90^{\circ}$ M--O--M bonds) are rather well studied theoretically and are considered in many publications (see, e.g.,~\cite{Gooden_book,KugKhoUFN82,Khomskii_book2014,StrKhomPRB2008pyroxenes}). There exists, however, the third, much less studied situation with neighboring MO$_6$ octahedra having a common face, i.e. having three common oxygens. This situation is illustrated schematically in Fig.~\ref{face_sharing}. Here, the superexchange occurs via three oxygens, with M--O--M angle for ideal (undistorted) MO$_6$ octahedra equal to about $70.5^{\circ}$. Note also, that in this case (as well as in the case of edge-sharing octahedra, Fig.~\ref{edge_sharing}) the direct $d$-$d$ hopping can be rather large (the metal--metal distance in case of Figs.~\ref{edge_sharing}  and~\ref{face_sharing} is usually rather short -- sometimes even shorter than such distance in the corresponding metal!); for the common face the corresponding orbital of $a_{1g}$ symmetry has a form shown for one TM ion in Fig.~\ref{face_sharing}. The situation with the orbital ordering and the form of the resulting spin and orbital exchange for compounds with face-sharing octahedra until recently was not known. Nevertheless, experimentally there are many TM compounds with such a geometry. These are, for example, hexagonal crystals like BaCoO$_3$~\cite{YamauraJSStCh1999BaCoO3}, CsCuCl$_3$~\cite{HirotsuJPhC1977CsCuCl3} or Ba$_9$Rh$_8$O$_{24}$\cite{Stitzer2001}, see right part of Fig.~\ref{face-sharing_arrays}, containing infinite columns of face-sharing ML$_6$ octahedra (L stands here for ligands O, Cl, ...), i.e. infinite stacking of pairs like that shown in Fig.~\ref{anti-ferro-orbit}(c). Many other similar systems have finite face-sharing blocks, e.g. Ba$_5$AlIr$_2$O$_{11}$~\cite{Terzic2015}, BaIrO$_3$~\cite{SiegristJLCM1991BaIrO3} or BaRuO$_3$~\cite{HongJSStCh1997BaRuO3,ZhaoJSStCh2007BaRuO3}, Ba$_3$CuSb$_2$O$_9$~\cite{KohlReinenZanorgAlChem1977Ba3CuSb2O9}. There can exist a more complicated connections of such blocks, like in Ba$_4$Ru$_3$O$_{10}$~\cite{CarimJSStCh2000Ba4Ru3O10}, middle part of Fig.~\ref{face-sharing_arrays}, with alternating 3--1 blocks, etc. Such systems have very diverse properties: some of them are metallic, but there are also good insulators among them, with very different magnetic properties. However, in any case, the first problem to consider for such systems is the question of a possible orbital and magnetic exchange in this geometry. This question was recently addressed in the paper \cite{MyPRB2015_SU4}; below we, first, shortly reproduce these results, and then we generalize them for certain specific situations, in particular for the case of different distortions of octahedra and for the strong spin--orbit coupling.

We will consider the form of the spin--orbital (``Kugel--Khomskii'') superexchange for TM with double or triple orbital degeneracy for neighboring TM ions with face-sharing octahedra. One surprising result of our study is that, whereas for doubly degenerate system of perovskite type with 180$^{\circ}$ M--O--M bonds, the form of orbital term in  Hamiltonian~\eqref{KK_model} is rather complicated~\cite{KugKhoUFN82}, for common face the situation similar to the symmetric model of Eq.~\eqref{symm_model} is realized. In effect, in such real systems the dominant term $\propto t^2/U$  in the spin-orbital superexchange has a very high symmetry, SU(4), i.e. such  materials may be real examples of applicability of such a fancy model (the higher order terms in this Hamiltonian, containing Hund's rule coupling, $J_H/U$ or $J_H/\Delta_{CT}$~\cite{ZSA_PRL1985chtrgap}, have a more complicated form, see below).

Often in such geometry the MO$_6$ octahedra have trigonal distortions (e.g., they are elongated or compressed along the vertical $z$ axis connecting TM ions in the chain). Such local distortions lead to splitting of $t_{2g}$ orbitals into an $a_{1g}$ singlet and $e_g^{\pi}$ doublet, Fig.~\ref{FigStruct}(b); the original $e_g$ ($e_g^{\sigma}$) doublet remains unsplit. Also the crystal structure itself leads to such trigonal crystal field splitting even for ideal undistorted MO$_6$ octahedra, in particular due to interactions with the transition metal ions in neighboring octahedra. One can show that if we have partially filled $e_g^{\pi}$ doublet, the resulting superexchange is very similar to the case of ``real'' $e_g$ electrons. Nevertheless,  there is one important difference when we are dealing with $t_{2g}$ states, in contrast to $e_g$ ones: whereas for $e_g$ states the orbital moment is quenched and the real relativistic spin--orbit coupling $\lambda\mathbf{lS}$ does not work in the first order, for $t_{2g}$ electrons it is not the case, and the SOC has to be taken into account. It can modify the resulting form of superexchange even when SOC is relatively weak. The effect of SOC can be especially important for heavy $4d$ and especially $5d$ elements, for which  $\lambda$ may be comparable with the Hund's rule coupling constant $J_H$ and even with the Hubbard interaction $U$. In this case, one may again need to go over to the description in terms of the effective total moment $\mathbf{j}=\mathbf{l} +\mathbf{S}$. As mentioned in Section II  for example for Ir$^{4+}$ ($t_{2g}^5$) the resulting picture would correspond to the doublet $j=1/2$.

The form of the resulting exchange for these effective Kramers doublets $j=1/2$ for the cases of common corner and common edge was presented in Sections II and III. How this interaction would look like for the face-sharing octahedra, e.g., in BaIrO$_3$, was not studied yet; we consider this case too and derive corresponding form of the superexchange. It turns out that for the face-sharing octahedra, the form of the exchange for this doublet is again more symmetric and has the Heisenberg form ${\bm\sigma}_i{\bm\sigma}_j$, i.e. it reminds the case of the $180^{\circ}$  Ir--O--Ir bonds.

\subsection{The model}

\begin{figure}[t]
\centering
\includegraphics[width=0.7\columnwidth]{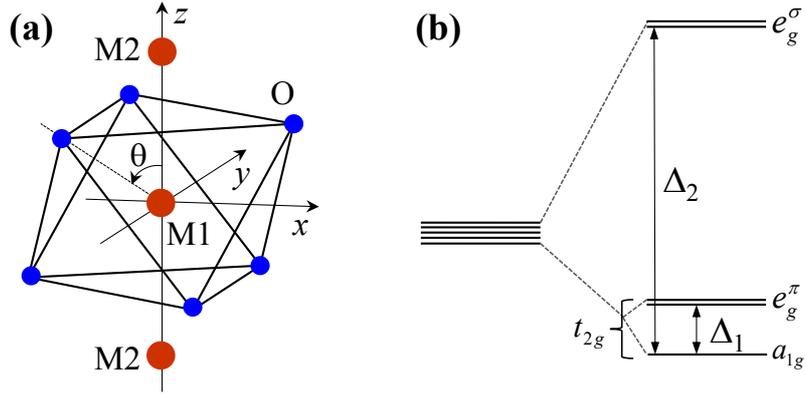}
\caption{(Color online) (a) Magnetic atom (M) surrounded by trigonally distorted oxygen (O) octahedron in transition metal compounds with face-sharing. Distortions are determined by the angle $\theta$; the value $\cos\theta_0=1/\sqrt{3}$ corresponds to undistorted octahedron. Magnetic atoms form a quasi-one-dimensional chain directed along the $z$ axis. (b) Crystal field splitting of $d$-orbitals of the magnetic atom. The splitting of $t_{2g}$ levels ($\Delta_1$) is due to both the trigonal distortions of oxygen octahedra and contribution from neighboring M atoms to the crystal field. The sign of $\Delta_1$ can be different depending on the type of distortions.}\label{FigStruct}
\end{figure}

We consider a linear chain of $3d$ ions located at the centers of  anion octahedra with face-sharing geometry. Two reference systems are of interest: the local tetragonal reference system of each magnetic ion and a global trigonal system, in which the $z$ axis is directed along the chain and the $x$ and $y$ axes are in the plane perpendicular to the chain. Notice that two nearest-neighbor ions have relatively rotated local axes and therefore we can not use the same tetragonal reference system for them. The local tetragonal reference systems are chosen such that the trigonal $z$ axis corresponds to the same $[111]$ direction for them.  In the further analysis, we choose the trigonal reference frame as shown in Fig.~\ref{FigStruct}(a). The trigonal distortions can be characterized by the angle $\theta$ also shown in this figure. For the ideal MO$_6$ octahedron, we have $\theta=\theta_0\equiv\arccos(1/\sqrt{3})$ = 54.74$^{\circ}$, while the M--O--M angle is $\beta_0=\pi-2\theta_0\approx70.5^{\circ}$.

The crystal field felt by the magnetic ions has an important component of cubic symmetry due to the octahedra of anions, and also a component with trigonal symmetry due to both the ions along the chain and the trigonal distortions of octahedra. In a octahedral field, the electron $d$ levels are split into a triple degenerate level ($t_{2g}$) and a doubly degenerate level ($e_g$). These levels can be further split by the trigonal field and the spin-orbit coupling. We study both these cases separately.

\subsection{Undistorted octahedra, $e_g$ levels}

As a minimum model for the chain, we take the Hubbard model for the case of degenerate electrons, in which we also made a simplifying assumption that the on-site Hubbard repulsion $U$ is the same for all orbitals (i.e. we effectively put the Hund's rule coupling $J_H$ to zero). This simplification is sufficient for our main purposes; we will mention possible modifications due to inclusion of $J_H$ when necessary. This Hamiltonian has the form
\begin{equation}\label{HHubbard}
H=\sum_{\langle ij\rangle}\sum_{\gamma\gamma'}\sum_{\sigma\sigma'}t_{ij}^{\gamma\gamma'}
c^{\dagger}_{i\gamma\sigma}c_{j\gamma'\sigma'}+
\frac{U}{2}\sum_{i}\sum_{\gamma\gamma'}\sum_{\sigma\sigma'}n_{i\gamma\sigma}n_{i\gamma'\sigma'}(1-\delta_{\gamma\gamma'}\delta_{\sigma\sigma'})\,,
\end{equation}
$c^{\dagger}_{i\gamma\sigma}$ and $c_{i\gamma\sigma}$ are the creation and annihilation operators of a $d$ electron with the orbital state $\gamma$ and spin projection $\sigma$ located at site $i$, $n_{i\gamma\sigma} = c^{\dagger}_{i\gamma\sigma}c_{i\gamma\sigma}$, and ${\langle ij\rangle}$ denotes the summation over the nearest-neighbor sites. The first term describes the kinetic energy and the second one corresponds to the on-site Coulomb repulsion, which we treat as the largest parameter (i.e. we consider the case of strong Mott insulators with the orbital degeneracy). As is shown below, one can reduce to this form also the situation when the effective $d$-$d$ hopping occurs via ligands [in this case, via three oxygens, see in Fig.~\ref{FigStruct}(a)].

Let us first consider the chain build up by the ideal metal--oxygen octahedra. We start by calculating the orbitals of interest at each site. It is known (see, e.g., Ref.~\cite{AbragamBleaney1970}) that both the trigonal field and the spin--orbit coupling do not split the $e_g$ levels. In the trigonal coordinate system, the $e_g$ doublet for two neighboring magnetic ions along the chain can be written as~\cite{BatesJPhC1971,MyPRB2015_SU4}
\begin{eqnarray}
|d_1\rangle&=&\frac{1}{\sqrt{3}}| x^2-y^2\rangle-\sqrt{\frac{2}{3}}| xz\rangle\,,\nonumber\\
|e_1\rangle&=&-\frac{1}{\sqrt3}| xy\rangle-\sqrt{\frac{2}{3}}| yz\rangle\label{e_g_e1}
\end{eqnarray}
for an ion M$_1$, and
\begin{eqnarray}
|d_2\rangle&=&\frac{1}{\sqrt{3}}| x^2-y^2\rangle+\sqrt{\frac{2}{3}}| xz\rangle\,,\nonumber\\
|e_2\rangle&=&-\frac{1}{\sqrt3}| xy\rangle+\sqrt{\frac{2}{3}}| yz\rangle
\label{e_g_e2}
\end{eqnarray}
for the nearest-neighbor ion M$_2$.

Electron hopping amplitudes entering Hamiltonian \eqref{HHubbard}, have in our case two contributions, which can be of the same order of magnitude for this particular geometry: the direct hopping between two magnetic ions along the chain, $t_{\gamma\gamma'}^{d-d}$,  and the indirect (superexchange) hopping via the anions, $t_{\gamma\gamma'}^{viaA}$. We consider both situations separately.

It is easy to see that the direct $d$-$d$ hopping exists only between the same orbitals. The corresponding hopping integrals can be expressed through the Slater--Koster parameters~\cite{SlaterKosterPR1954}
\begin{equation}
\langle xy|\hat{t}|xy\rangle = \langle x^2-y^2|\hat{t}|x^2-y^2\rangle = V_{dd\delta}\,,
\end{equation}
\begin{equation}
\langle yz|\hat{t}|yz\rangle = \langle xz|\hat{t}|xz\rangle = V_{dd\pi}\,.
\end{equation}
So, in effect, we have only diagonal (and equal) hoppings
\begin{equation}
t^{d-d} = t^{d-d}_{|d_2\rangle|d_1\rangle}=t^{d-d}_{|e_2\rangle|e_1\rangle}=
\frac{1}{3}V_{dd\delta}+\frac{2}{3}V_{dd\pi}
\end{equation}
and
\begin{equation}
t^{d-d}_{|e_2\rangle|d_1\rangle}=t^{d-d}_{|e_2\rangle|d_1\rangle}=0\,.
\end{equation}

Similarly, one can show that the hopping integrals via intermediate oxygen ions, after we sum over all three of them, has the same feature~\cite{MyPRB2015_SU4}
\begin{equation}
t^{viaA}_{\gamma\gamma'}=t=t_0\delta_{\gamma\gamma'}\,,\;\;
t_0=\frac{3}{2}(t_1+t_2)\,,
\end{equation}
where $t_1$ and $t_2$ are the hopping integrals via one of the oxygen ions $ t_1 = \langle d_1|\hat{t}^{via O}|d_2\rangle$, $t_2 =\langle e_1|\hat{t}^{via O}|e_2\rangle$.  As a result, we have here exactly the same situation as in the symmetric model described in Section II, with hopping between two degenerate orbitals satisfying the relations $t_{11}=t_{22}=t$ and  $t_{12}=0$, so that actually in the main approximation we have SU(4) spin--orbital model \eqref{symm_model} also for the case of face-sharing transition metal compounds.

This is a rather general result based only on the existence of the three-fold trigonal axis and it does not depend on the specific features of the superexchange paths.

When one goes beyond the lowest order and includes the Hund's rule coupling $J_H$, the total exchange takes the form  (see also ~\cite{KugKhoUFN82})
\begin{eqnarray}
H_{eff}&=&\frac{t^2}{U}\sum_{\langle ij\rangle}\left\{\left(\frac{1}{2}+\mathbf{S}_i\mathbf{S}_j\right)
\left(\frac{1}{2}+{\bm\tau}_i{\bm\tau}_j\right)+\right.\nonumber\\
&&\frac{J_H/U}{1-(J_H/U)^2}\left[2\left({\bm\tau}_i{\bm\tau}_j-
\tau^z_i\tau^z_j\right)-\left(\frac{1}{2}+2\mathbf{S}_i\mathbf{S}_j\right)
\left(\frac{1}{2}-2\tau^z_i\tau^z_j\right)\right]+\label{KK_model1}\\
&&\left.\frac{(J_H/U)^2}{1-(J_H/U)^2}\left[-\left(\frac{1}{2}-2\tau^z_i\tau^z_j
\right)+2\left(\frac{1}{2}+\mathbf{S}_i\mathbf{S}_j\right)
\left({\bm\tau}_i{\bm\tau}_j-\tau^z_i\tau^z_j\right)\right]\right\}\,.\nonumber
\end{eqnarray}

Let us notice here, that, strictly speaking, similar to the case of common edge, Sec. III, Hamiltonian~\eqref{KK_model1} is valid only for the case of Mott--Hubbard insulators, when $U<\Delta_{CT}$. The opposite limit of charge-transfer insulators, $U>\Delta_{CT}$, requires separate analysis.

\subsection{Undistorted octahedra, $t_{2g}$ case}
The systems with TM ions surrounded by octahedra with common face typically have the trigonal symmetry, even for regular MO$_6$ octahedra. In reality, however, these octahedra are also usually distorted. Such distortions could preserve the $C_3$ symmetry and would correspond to a compression or stretching of these octahedra along the $z$ direction shown in Fig.~\ref{FigStruct}(a). The trigonal crystal field (CF) does not split $e_g$ levels, but leads to a splitting of $t_{2g}$ levels into an $a_{1g}$ singlet and $e_g^{\pi}$ doublet. Which levels lies lower, singlet or doublet, depends on the sign of trigonal CF.
In the trigonal coordinate system ($z$ axis along the chain), these wave functions have the form: for the $a_{1g}$ singlet
\begin{equation}
|a_1\rangle=|3z^2-r^2\rangle\,,\label{a1_singlet}
\end{equation}
and for the  $e_g^{\pi}$ doublet
\begin{eqnarray}
|b_1\rangle&=&-\frac{2}{\sqrt{6}}|xy\rangle+\frac{1}{\sqrt{3}}|yz\rangle\,,\nonumber\\
|c_1\rangle &=&\frac{2}{\sqrt{6}}|x^2-y^2\rangle+\frac{1}{\sqrt{3}}|xz\rangle\,,\label{t2g_b1}
\end{eqnarray}
for an ion M$_1$, and the same singlet
\begin{equation}
|a_2\rangle =|3z^2-r^2\rangle\,,\label{a2_singlet}
\end{equation}
and a doublet,
\begin{eqnarray}
|b_2\rangle &=&-\frac{2}{\sqrt{6}}|xy\rangle
-\frac{1}{\sqrt{3}}|yz\rangle\,,\nonumber\\
|c_2\rangle &=&\frac{2}{\sqrt{6}}|x^2-y^2\rangle
-\frac{1}{\sqrt{3}}|xz\rangle\,,\label{t2g_c2}
\end{eqnarray}
for the nearest-neighbor ion M$_2$.

These expressions are valid for the case of the ideal MO$_6$ octahedra, where M--O--M angle is about 70.5$^{\circ}$. The trigonal distortions lead to the modifications of the $e_g^{\pi}$ wave functions. A detailed description of such modifications is given in the Appendix. These modifications, however, do not change the main conclusion, see below,  that also here only diagonal hoppings are nonzero.

If one electron occupies the $a_{1g}$ level, there remains no orbital degeneracy, and the spin exchange would be trivially antiferromagnetic. More interesting is the case of one electron (or hole) at the $e_g^{\pi}$ level. One can show \cite{MyPRB2015_SU4} that in this case, similar to the case of ``real'' $e_g$ electrons discussed in subsection B, we have a symmetric model with the hoppings $\langle b_1|\hat{t}|b_2\rangle = \langle c_1|\hat{t}|c_2\rangle = t$ and $\langle b_1|\hat{t}|c_2\rangle = 0$, so, we also eventually have here the resulting spin--orbital model  \eqref{symm_model} with the SU(4) symmetry, or the more general exchange \eqref{KK_model1}. The detailed form of wave functions $|b\rangle$ and $|c\rangle$ depends on the noncubic (here trigonal) crystal field, and it is different from those written above (see Appendix), but it does not change qualitative conclusions and only changes numerical values of exchange constants in Hamiltonians \eqref{symm_model} and \eqref{KK_model1}.

\subsection{Role of spin--orbit interaction}

\begin{figure}[t]
\centering
\includegraphics[width=0.6\columnwidth]{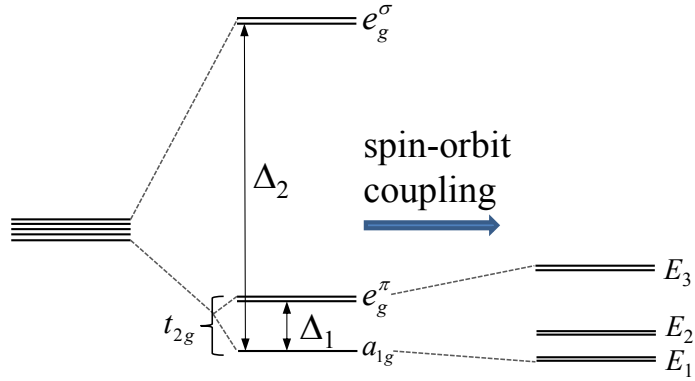}
\caption{(Color online) Spin-orbit interaction splits the $e_{g}^{\pi}$ and $a_{1g}$ energy levels into three Kramers doublets with energies $E_1$, $E_2$, and $E_3$. For undistorted octahedra and negligible effect of nearest-neighbor magnetic atoms there is one quartet with energy $E_1=E_2=-\lambda/2$ and one doublet with energy $E_3=\lambda$ (see the text).}\label{FigSO}
\end{figure}

We already presented above, in Sections II and III,  the form of the effective exchange interaction in the case of very strong spin--orbit coupling, for which, e.g. for ions like Ir$^{4+}$ or Ru$^{3+}$ with the $d^5$ configuration, one could sometimes reduce the whole description to that of a separate $j=1/2$ Kramers doublet (see $E_3$ doublet in Fig.~\ref{FigSO}). In this Subsection, we consider the same problem for the case of systems with common face, taking into account also trigonal crystal field, mentioned in the previous section.

First, we consider the simplest case of regular octahedra with the degenerate $t_{2g}$ orbital triplet for the same case of ions like Ir$^{4+}$ for strong spin--orbit coupling. For this case, the wave functions of the separate $j=1/2$ doublet $E_3$ for the site M1 are  analogous to  those discussed in \cite{Khomskii_book2014,JackKhal_PRL2009,AbragamBleaney1970},
\begin{eqnarray}
|j=+1/2\rangle &=& \frac{1}{\sqrt 3}\left[-|a_{1g},\uparrow\rangle + |(-c_1 + ib_1), \downarrow\rangle \right]\,, \\ \nonumber
|j=-1/2\rangle &=& \frac{1}{\sqrt 3} \left[-|a_{1g}, \downarrow\rangle + |(c_1 + i b_1),\uparrow\rangle\right]\,,
\label{j=1/2}
\end{eqnarray}
and similar expression for the site M2 with the wave functions $|b_2\rangle, |c_2\rangle$ instead of $|b_1\rangle, |c_1\rangle$, see \eqref{a1_singlet}--\eqref{t2g_c2}. Here, $a_{1g}$ states correspond to $l_{eff}^z =0$, and the states $|b_1 \pm i c_1\rangle$ -- to states with $l_{eff}^z = \pm 1$ for quantization along the $z$ axis in the global coordinate system of Fig.~\ref{FigStruct}(a).

Projecting into this manifold, we easily obtain, using  wave functions \eqref{a1_singlet}--\eqref{t2g_c2}, that the hopping matrix elements are diagonal and equal to each other both for direct $d$-$d$ hopping and hopping via oxygens,
\begin{equation} \label{hopping}
 \langle 1/2|\hat{t}|1/2\rangle = \langle -1/2|\hat{t}|-1/2\rangle = t, \quad   \langle1/2|\hat{t}|-1/2\rangle = 0\,.
\end{equation}
Similar to the case of Section II, we immediately  see that in this case the exchange written in terms of effective spin 1/2 of $j = 1/2$ doublet again takes the form of a Heisenberg interaction \eqref{JackKhal}.

Let us now consider, which modifications can we expect in the case of trigonal splitting of $t_{2g}$ states, and also what one could have for other electron occupations. Note that the treatment below deals with one-electron levels,  i.e. it effectively corresponds not to $LS$ (Russel--Sounders), but to $JJ$ coupling. This is actually the assumption implicitly made in most treatment of systems with strong SOC like iridates, although it is not always  stated explicitly.

As mentioned above, for systems with common face, the trigonal splitting of $t_{2g}$ levels is very typical. The detailed treatment of this situation is given in Appendix. Here, we summarize and qualitatively explain the main findings.  As can be shown, inclusion of both (strong) SOC and trigonal CF leads to the structure of levels shown in Fig.~\ref{FigSO}. Typically, except some isolated points in parameter space, $t_{2g}$ levels are split into three Kramers doublets (for the cubic CF, the doublets $E_1$ and $E_2$ in Fig.~\ref{FigSO} are degenerate and combine into a $j=3/2$ quartet). If we have e.g. the system like Ir$^{4+}$ ($d^5$), we would have one electron at the $E_3$ doublet. In general, the wave functions \eqref{j=1/2} of this doublet would be different from those of unsplit $t_{2g}$ triplet. Nevertheless, at least for large $t_{2g}$--$e_g$ splitting 10$Dq$ (ignoring possible admixture of ``real'' $e_g$ states), these would be all composed of the superposition of  $t_{2g}$ functions \eqref{a1_singlet}--\eqref{t2g_c2} or of functions with $|l_z=0\rangle = |a_{1g}\rangle$ and $|l^z =\pm 1\rangle = (1/\sqrt2)|b_{1,2} \pm i c_{1,2}\rangle$, in the form
\begin{eqnarray}
|+\rangle &=& c_0 |l^z=0, \uparrow\rangle + c_1 |l^z = +1, \downarrow\rangle, \\ \nonumber
|-\rangle &=& c_0 |l^z=0, \downarrow\rangle + c_{-1} |l^z = -1, \uparrow\rangle\,,
\label{l_wavefunc}
\end{eqnarray}
with $c_1 = c_{-1}$  and  $c_0^2+c_1^2 = 1$.  Again, in general case, when we take into account that hopping matrix elements are nonzero only for diagonal hopping, $\langle0|\hat{t}|0\rangle = t_0,  \langle+|\hat{t}|+\rangle = \langle-|\hat{t}|-\rangle = t_1$, and  non-diagonal hoppings are zero, $\langle0|\hat{t}|\pm1\rangle = \langle+|\hat{t}|-\rangle = 0$, we finally obtain that also in this general case, we have the same symmetric situation, with $\langle+|\hat{t}|+\rangle = \langle-|\hat{t}|-\rangle = t, \langle+|\hat{t}|-\rangle = 0$. i.e. we have only a diagonal and equal hopping within the $E_3$ doublet.

Consequently, for this case, we would also get in general in the main approximation the simple Heisenberg interaction (3) for this Kramers doublet (if indeed SOC is strong enough so that this doublet is well separated from the $E_1$ and $E_2$ levels). This can be traced back to the fact that electron hopping, on the one hand, preserves  spin, $t_{\uparrow,\uparrow} = t_{\downarrow,\downarrow} = t, t_{\uparrow,\downarrow} = 0$, but also conserves the orbital moment, so that $\langle0|\hat{t}|0\rangle,  \langle+1|\hat{t}|+1\rangle$, and $\langle-1|\hat{t}|-1\rangle$ are nonzero, but hoppings with the change of orbital moment disappear, i.e. nondiagonal matrix elements are zero. This in effect is responsible for the realization of the symmetric model, which finally gives for very strong SOC (isolated doublet $E_3$) the Heisenberg interaction. However, in the general situation, for arbitrary relation between SOC and CF splitting and for other signs of this trigonal CF  and  other filling of $d$ levels, the situation could be very different and would require special treatment (some basis for which is presented in Appendix).

\section{Conclusions}\label{concl}

In this paper, we presented a survey of the spin--orbital interaction for orbitally-degenerate Mott insulators for different local geometries (MO$_6$ octahedra with common corner, common edge and common face), paying main attention to the ``third case'' of the common face, which, strangely enough, was practically not considered yet in the existing literature. The main message is that the general form and the details of the spin--orbital (``Kugel--Khomskii'')  exchange interaction very strongly depends on this local geometry, so that the commonly accepted paradigm  (ferro-orbitals $\leftrightarrow$ antiferro-spins, and {\it vice versa}, antiferro-orbitals $\leftrightarrow$ ferro-spins), derived for 180$^{\circ}$ degree metal-oxygen-metal bonds (common corner), is not valid in general.

Rather surprisingly, the ``third case'' of octahedra with common face  turns out to be in some sense simpler and more symmetric than the other two situations, despite apparently more complicated local geometry (exchange via three oxygens, with ``not simple'' M--O--M angle about 70$^{\circ}$, etc.).  In particular, for doubly-degenerate case ($e_g$ orbitals or $e_g^{\pi}$ doublet produced out of $t_{2g}$ triplet by trigonal splitting, typical for this case) the effective spin--orbital model has a highly symmetric form \eqref{symm_model}, i.e. it contains scalar products of both spin operators $\bm S$ and orbital operators $\bm \tau$ describing the orbital doublet. The resulting exchange \eqref{symm_model}  has not only  SU(2)$\times$SU(2) symmetry dictated by these scalar products, but they enter with such coefficients that the resulting symmetry is even much higher -- it is SU(4),  the very nice theoretical model, which for example is exactly soluble in the 1D case, etc.

Thus the materials with MO$_6$ octahedra sharing common face can be a good model systems to study possible manifestation of this high symmetry. Similarly, local geometry largely determines the resulting form of the exchange in case of very strong spin-orbit coupling $\lambda\mathbf{LS}$ -- the situation typical for 4$d$ and 5$d$ systems. As is already known, for example for ions with $d^5$ configuration, such as popular nowadays Ir$^{4+}$, the exchange Hamiltonian for the lowest Kramers doublet $j=1/2$ has the Heisenberg form for 180$^{\circ}$ M--O--M bonds (common corner), but it is highly anisotropic (locally Ising) interaction for 90$^{\circ}$ bonds (common edge). Again, the situation for systems  with the common face turns out to be simpler in case of strong spin--orbit coupling too. The exchange for Kramers doublets $j=1/2$ again has the Heisenberg form, $H \sim J \sum{{\bm \sigma}_i{\bm \sigma}_j}$, where ${\bm \sigma}$ is the effective spin describing the $j=1/2$ doublet.

We have also shown that the account taken of the trigonal splitting, very typical for the case of common face, does not change the situation qualitatively, although there appear definite quantitative changes.

These situations considered above, although the most typical ones, do not exhaust all the variety of local geometries met in transition metal compounds. Thus, TM ions can be not in O$_6$ octahedra, but for example in O$_4$ tetrahedra, like A sites in spinels or Co ions in YBaCo$_4$O$_7$. Or they may be in trigonal bipyramids (Mn in YMnO$_3$) or in prisms (half of Co ions in Ca$_3$Co$_2$O$_6$), etc. Every such case requires special treatment, one cannot uncritically transfer the know-how acquired in considering spin--orbital system in, say, perovskites to these cases. Important conclusion  is that this concerns not only these, more complicated cases, but also the situation with more conventional materials containing TM ions in O$_6$ octahedra, may be more complicated than usually assumed. Apart from the specific results for the ``third case'' of octahedra sharing common face, this warning is the main message of the present paper.

\begin{acknowledgments}
We highly appreciate an opportunity to publish this paper in the special issue of JETP dedicated to the 85th birthday of L.V. Keldysh. We gratefully acknowledge a long-term inspiring influence of his ideas on our work.

This work is supported by the Russian Foundation for Basic Research (projects 14-02-00276-a, 14-02-0058-a, 13-02-00050-a, 13-02-00909-a, and 13-02-00374-a), by the Russian Science Support Foundation, by the Russian Federal Agency for Scientific Organizations (theme ``Electron'' No. 01201463326 and program 02.A03.21.0006), by the Ural Branch of Russian Academy of Sciences, by the German projects DFG GR 1484/2-1 and FOR 1346, and by K\"oln University via the German Excellence Initiative.
\end{acknowledgments}

\appendix* \section{Effects of trigonal distortions}

\subsection{Face-sharing geometry with trigonal distortions: Wave functions and energy levels}

Actually, we never deal with ideal octahedra. In particular, the chain of face-sharing octahedra is usually stretched or compressed. The effect of such distortions can be described in terms of the crystal field of trigonal symmetry. Below in Appendix 1 for completeness we shortly reproduce and extend the results of the treatment of trigonal splitting, carried out in Ref. \cite{MyPRB2015_SU4}; these results would be important for us for the general treatment of the effect of the SOC in Appendix 2.

Elementary building block of the transition metal compounds with face-sharing octahedra is shown in Fig.~\ref{FigStruct}(a). Each magnetic atom is surrounded by a distorted oxygen octahedron. Distortions can be described by single parameter $\theta$, which is the angle between $z$-axis and the line connecting M and O atoms [see Fig.~\ref{FigStruct}(a)]. For undistorted octahedron, we have $\theta=\theta_0=\arccos(1/\sqrt{3})$. Crystal field splits $5$-fold degenerate $d$ electron levels of the transition metal atom into two doubly degenerate $e_{g}^{\sigma}$, $e_{g}^{\pi}$ levels, and $a_{1g}$ level, like shown in Fig.~\ref{FigStruct}(b). The energy difference $\Delta_1$ between $e_{g}^{\pi}$ and $a_{1g}$ levels can be positive or negative depending on the type of trigonal distortions and other parameters of the system.

We should find the wave functions of the $e_{g}^{\sigma}$, $e_{g}^{\pi}$, and $a_{1g}$ levels in the case of distorted octahedra. We neglect first the contribution to the crystal field from a neighboring magnetic cations. In the point-charge approximation, the crystal field potential acting onto a chosen cation located at point $\mathbf r$ can be represented as a sum of Coulomb terms
\begin{equation}
V(\mathbf r) = v_0\sum_i\frac{r_0}{|{\mathbf r} - {\mathbf r}_i|}\, ,
\end{equation}
where ${\mathbf r}_i$ are the positions of ligand ions. For $d$ states, the existence of the three-fold symmetry axis leads to a significant simplification of the expression for the crystal field
\begin{equation}\label{VvsR}
V({\mathbf r}) = v_0(r) + v_1(r)\sum_{s=1}^{3} P_2(\cos\theta_s)+v_2(r)\sum_{s=1}^{3} P_4(\cos\theta_s)\,,
\end{equation}
where $P_2$ and $P_4$ are the Legendre polynomials, $P_2(x) = \frac{1}{2}(3x^2- 1)$ and $P_4(x) = \frac{1}{8}(34x^4- 30x^2 + 3)$. Here, we took into account  the symmetry in the arrangement of two opposite edges of the ligand octahedron and as a result, we have
\begin{equation}
\cos\theta_s = \cos\theta \cos\theta' + \sin\theta \sin\theta' \cos \left( \phi' - \frac{2\pi s}{3}\right)\,,
\end{equation}
where $\theta'$ and $\phi'$ describe the direction of $\mathbf{r}=r\{\sin\theta'\cos\phi',\,\sin\theta'\sin\phi',\,\cos\theta'\}$.

Then, it is necessary to find the matrix elements of the crystal field for the complete set of $d$ functions and to diagonalize the corresponding matrix. This gives us both the wave functions of $e_{g}^{\sigma}$, $e_{g}^{\pi}$, and $a_{1g}$ levels and their energies, depending on the trigonal distortions, the details of such calculations can be found in \cite{MyPRB2015_SU4}. Choosing the reference frame like shown in Fig.~\ref{FigStruct}(a), we eventually obtain for the wave functions the expressions having the forms similar to those discussed above for the case of undistorted octahedra. Thus, for $e_g$ levels ($e_g^{\sigma}$ orbitals) we have [cf. Eqs.~\eqref{e_g_e1}--\eqref{e_g_e2}]
\begin{eqnarray}
|d_{1,2}\rangle&=&\sin\!\frac{\alpha}{2}\,|x^2-y^2\rangle \mp\cos\!\frac{\alpha}{2}\,|xz\rangle\,,\nonumber\\
|e_{1,2}\rangle&=&-\sin\!\frac{\alpha}{2}\,|xy\rangle \mp\cos\!\frac{\alpha}{2}\,|yz\rangle\,.\label{e_g_e12_gen}
\end{eqnarray}
For $t_{2g}$ orbitals we have the same $a_{1g}$ singlet, Eqs.~\eqref{a1_singlet} and \eqref{a2_singlet}, and the $e_g^{\pi}$ doublet [cf. with Eqs.~\eqref{t2g_b1} and \eqref{t2g_c2}]
\begin{eqnarray}
|b_{1,2}\rangle&=&-\cos\!\frac{\alpha}{2}\,|xy\rangle \pm\sin\!\frac{\alpha}{2}\,|yz\rangle\,,\nonumber\\ |c_{1,2}\rangle&=&\cos\!\frac{\alpha}{2}\,|x^2-y^2\rangle \pm\sin\!\frac{\alpha}{2}\,|xz\rangle\,.\label{t2g_c12_gen}
\end{eqnarray}
The $\mp$ and $\pm$ signs in the above expressions for cation wave functions for neighboring magnetic ions occur since the oxygen octahedra surrounding neighboring metal ions are transformed to each other by the $180^{\circ}$ rotation about $z$ axis. Parameter $\alpha$ in Eqs.~\eqref{e_g_e12_gen} and~\eqref{t2g_c12_gen} depends on the trigonal distortions as
\begin{equation}\label{alpha}
\cos\alpha=\frac{a}{\sqrt{a^2+b^2}},\,\,\,a=a_2+a_4\, ,
\end{equation}
where \begin{eqnarray}
a_4&=&-{3\over 2}\left({5\over 2}\cos^4\theta -
{15\over 7}\cos^2\theta + {3\over 14}\right)\,,\nonumber\\
a_2&=&{27\over 35}\kappa\left(3\cos^2\theta -1\right)\,,
\label{CF_parameters}\\
b&=&3\sin^3\theta\cos\theta\,.\nonumber
\end{eqnarray}
Parameter $\kappa$ is defined as
\begin{equation}
\kappa=r_0^2\frac{\int\limits_0^{\infty}v_1(r)R_d^2(r)r^2dr}
{\int\limits_0^{\infty}v_2(r)R_d^2(r)r^2dr}\,,
\end{equation}
where $r_0$ is the cation--ligand distance. Parameter $\kappa$ depends also on the crystal fields $v_{1,2}(r)$ [see the expansion~\eqref{VvsR}] and the radial part of the wave function $R_d(r)$ for $d$ electrons.  The value of $\kappa$ depends on the material under study. Both semianalytical and {\it ab initio} calculations done in Ref.~\cite{MyPRB2015_SU4} give the estimate $\kappa\sim0.1$--$1$.

For the ideal octahedron, we have $\alpha=\alpha_0\equiv\pi-2\theta_0=\arccos(1/3)$. Substituting this value to Eqs.~\eqref{e_g_e12_gen} and~\eqref{t2g_c12_gen}, we arrive at the results presented in Sec. IVc. Stretching (compression) of oxygen octahedra tends to make $\alpha<\alpha_0$ ($\alpha>\alpha_0$).

\begin{figure}[t]
\centering
\includegraphics[width=0.32\columnwidth]{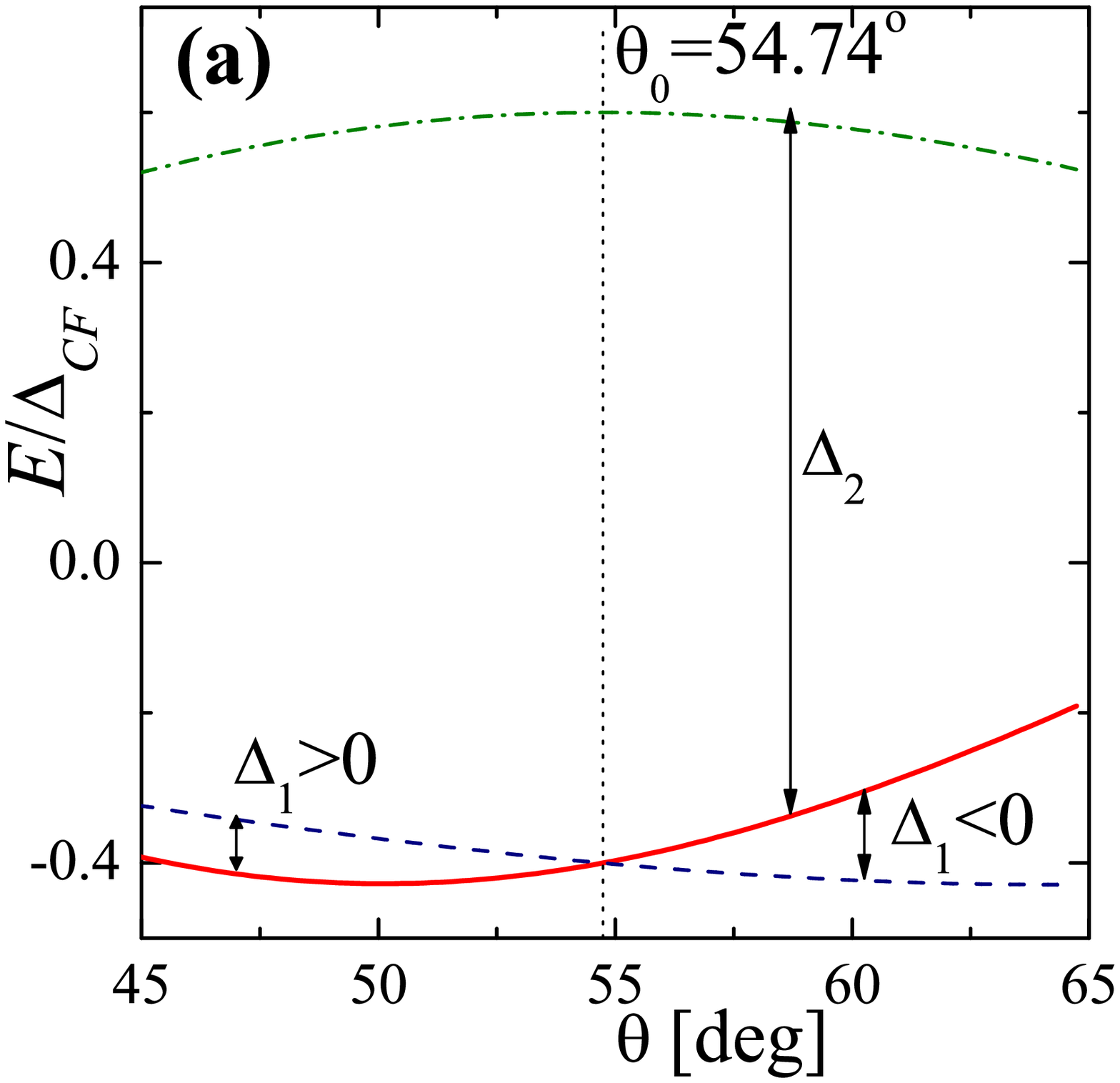}
\includegraphics[width=0.32\columnwidth]{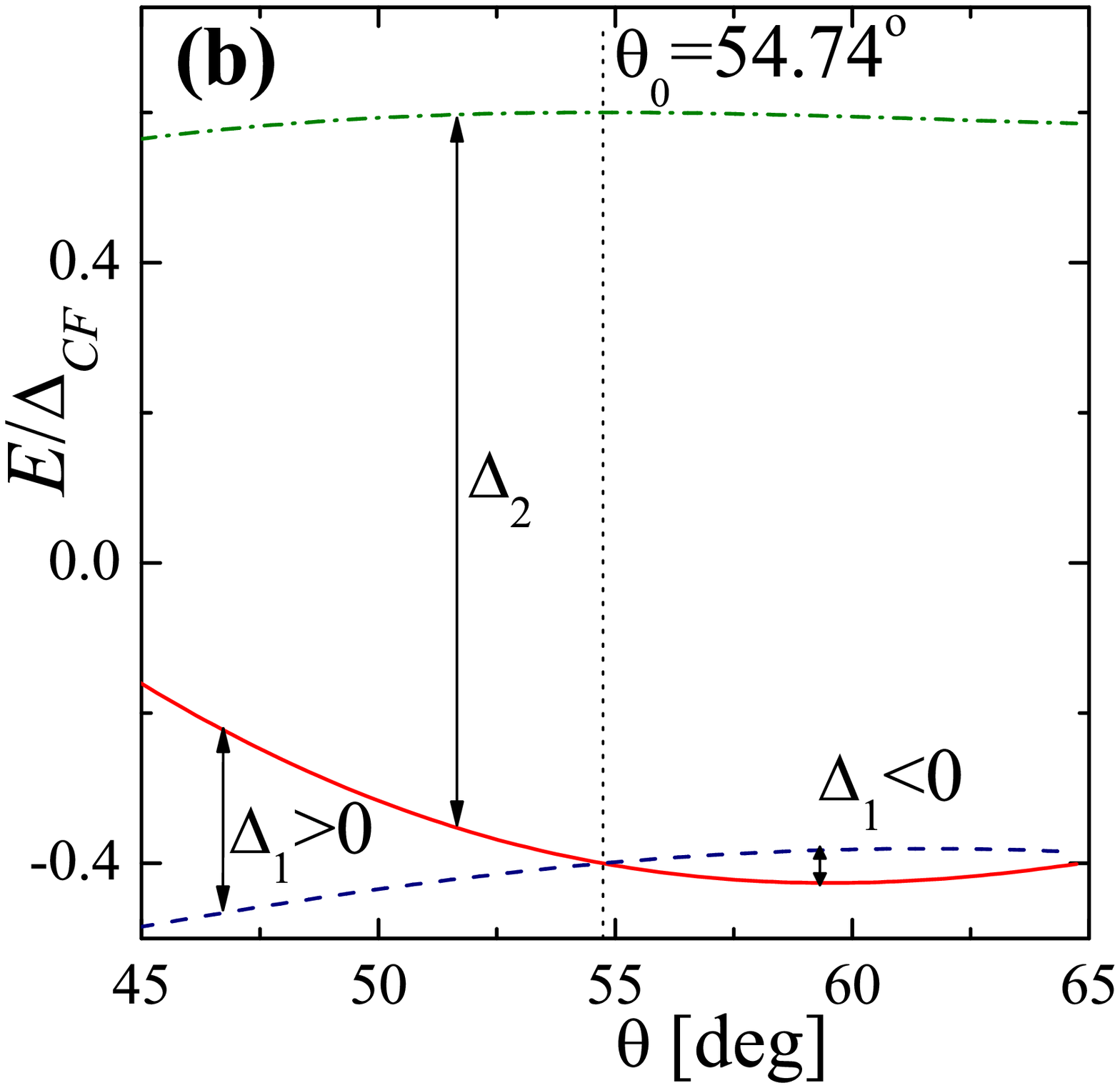}
\includegraphics[width=0.32\columnwidth]{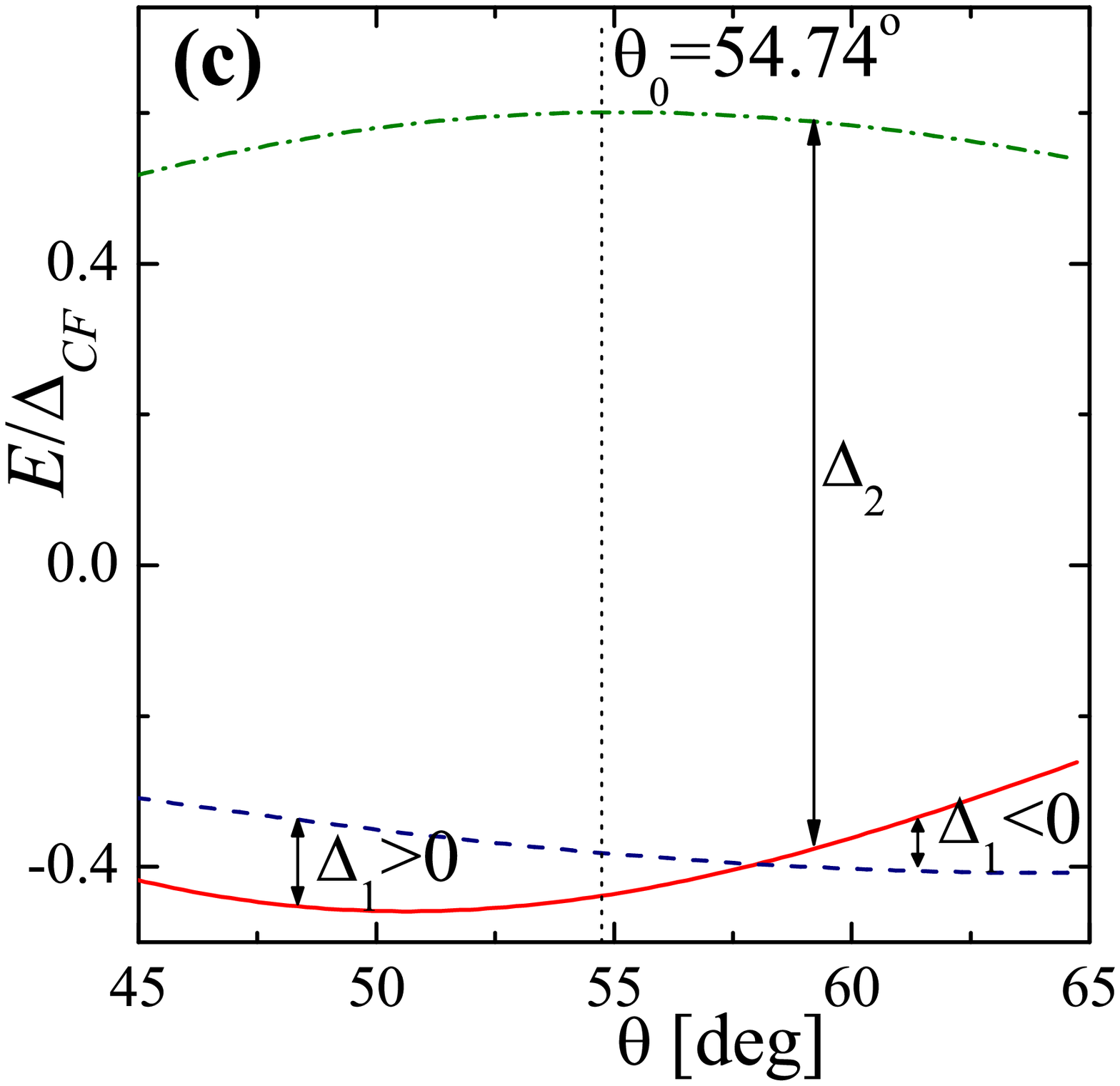}
\caption{(Color online) Energies $E_{a_{1g}}$ (red solid line),  $E_{e_{g}^{\pi}}$ (blue dashed line), and $E_{e_{g}^{\sigma}}$ (green dot-dash line) versus angle $\theta$ calculated for $\kappa=0.1$, $Z^{*}=0$ (panel {\it a}),  $\kappa=1$, $Z^{*}=0$ (panel {\it b}), and $\kappa=0.1$, $Z^{*}=3$ (panel {\it c}). For definition of $\kappa$ and $Z^{*}$, see the text.}\label{FigEvsTheta}
\end{figure}

Let us analyze now the behavior of $e_{g}^{\sigma}$, $e_{g}^{\pi}$, and $a_{1g}$ energy levels on the trigonal distortions. Figure~\ref{FigEvsTheta} shows the dependence of the energies $E_{e_{g}^{\sigma}}$, $E_{e_{g}^{\pi}}$, and $E_{a_{1g}}$ on the angle $\theta$ varying near $\theta_0\cong54.74^{\circ}$ calculated for two different values of $\kappa$. For ideal octahedron, we have $\Delta_1=E_{e_{g}^{\pi}}-E_{a_{1g}}=0$ for any $\kappa$. The sign of $\Delta_1$ depends both on the type of trigonal distortion (stretching for $\theta<\theta_0$, or compression for $\theta>\theta_0$) and the value of $\kappa$. When $\kappa\lesssim0.5$, the stretching (compression) of octahedron makes $\Delta_1>0$ ($\Delta_1<0$), while for  $\kappa\gtrsim0.5$ the situation is opposite [see Fig.~\ref{FigEvsTheta}(a,b)]. Note, that we consider only distortions with $\theta$ near $\theta_0$, when $\Delta_2=E_{e_{g}^{\sigma}}-E_{a_{1g}}\sim10Dq\gg|\Delta_1|$. Such a situation corresponds to the experiment for all known compounds. At the same time, the sign of $\Delta_1$ can be different for different systems. For example,  $\Delta_1>0$ for the BaCoO$_3$ with the chains of face-sharing Co$^{4+}$O$_6$ octahedra~\cite{YamauraJSStCh1999BaCoO3}.

These results were obtained neglecting the effect of neighboring metal atoms in the chain. Taking into account the contribution to the crystal field from these atoms modifies the parameter $a_2$ in the following manner
\begin{equation}
a_2\to a_2-\frac{27\kappa}{35}\frac{Z^*}{12\cos^2\theta}\,,
\end{equation}
where $Z^{*}$ is the effective charge (in units of $e$) of the metal ion. Parameters $a_4$ and $b$, as well as the relations~\eqref{e_g_e12_gen}--\eqref{alpha} remain the same. The crystal field from neighboring metal atoms tends to increase $\Delta_1$ and parameter $\alpha$. The dependence of the $e_{g}^{\sigma}$, $e_{g}^{\pi}$, and $a_{1g}$ energies levels  calculated for non-zero $Z^*$ is shown in Fig.~\ref{FigEvsTheta}(c).

The wave functions Eqs.~\eqref{e_g_e12_gen} and~\eqref{t2g_c12_gen} are the generalization of those considered in the previous section to the case of arbitrary trigonal distortion characterized by an angle $\alpha$. It is quite straightforward to demonstrate that the similar structure of $e_{g}^{\sigma}$ and  $e_{g}^{\pi}$ wave function leads to the same symmetric spin--orbital Hamiltonian~\eqref{symm_model} with the SU(4) symmetry at any given value of $\alpha$.

\subsection{Spin--orbit coupling in the case of trigonally distorted octahedra}

We start from the analysis of the structure of $d$-electron levels. The spin--orbit interaction Hamiltonian has a form
\begin{equation}\label{H_SO}
H_{\text{SO}}=\lambda\mathbf{l}\mathbf{S}\,.
\end{equation}
To find eigenenergies and eigenfunctions of the $d$-electron levels, we represent the orbital momenta operators $l_z$ and $l_{\pm}=l_{x}\pm il_{y}$ in the basis of wave functions $|\mu\rangle$, in which the matrix describing the crystal field splitting is diagonal
\begin{equation}
\left(\hat{\tilde{l}}_{z}\right)_{\mu\nu}=\langle\mu|\,l_z\,|\nu\rangle,\;\;
\left(\hat{\tilde{l}}_{\pm}\right)_{\mu\nu}=\langle\mu|\,l_{\pm}\,|\nu\rangle\,.
\end{equation}
The basic wave functions are $|\mu\rangle=\{|e_{g1}^{\sigma}\rangle,\,|e_{g2}^{\sigma}\rangle,
\,|e_{g1}^{\pi}\rangle,|e_{g2}^{\pi}\rangle,\,|a_{1g}\rangle\}$, where
\begin{equation}
|e_{g1}^{\sigma}\rangle=|d_1\rangle\,,\;|e_{g2}^{\sigma}\rangle=
|e_1\rangle\,,\;|e_{g1}^{\pi}\rangle=|b_1\rangle\,,\;
|e_{g2}^{\pi}\rangle=|c_1\rangle\,,
\end{equation}
for magnetic ion M$_1$, and
\begin{equation}
|e_{g1}^{\sigma}\rangle=|d_2\rangle\,,\;|e_{g2}^{\sigma}\rangle=
|e_2\rangle\,,\;|e_{g1}^{\pi}\rangle=|b_2\rangle\,,\;
|e_{g2}^{\pi}\rangle=|c_2\rangle\,,
\end{equation}
for magnetic ion M$_2$ [for definition of $|b_{1,2}\rangle$, $|c_{1,2}\rangle$, $|d_{1,2}\rangle$, and $|e_{1,2}\rangle$ see Eqs.~\eqref{e_g_e12_gen} and~\eqref{t2g_c12_gen}]. The basic wave function $|a_{1g}\rangle$ is the same for both magnetic ions; it is given by Eq.~\eqref{a1_singlet}. In this basis, the $5\times5$ matrix $\hat{V}$, describing the crystal field, is diagonal and has a form $\hat{V}=\diag\{\Delta_2,\Delta_2,\Delta_1,\Delta_1,0\}$. The spin--orbit coupling breaks the degeneracy of $d$ levels on the electron spin. Keeping this in mind, we introduce the second (spin) index to the basic wave functions $|\mu\rangle\to|\mu,\sigma\rangle$ with $\sigma=\uparrow,\,\downarrow$. The total Hamiltonian, then, can be presented in the form of the $10\times10$ matrix, which can be written in the following block-matrix form
\begin{equation}
\hat{H}=\left(\begin{array}{cc}\hat{V}&0\\0&\hat{V}\end{array}\right)+
\frac{\lambda}{2}\left(\begin{array}{cc}\hat{\tilde{l}}_{z}&\hat{\tilde{l}}_{-}\\
\hat{\tilde{l}}_{+}&-\hat{\tilde{l}}_{z}\end{array}\right)\,.
\end{equation}
Diagonalization of $\hat{H}$ gives us the structure of electron levels in the presence of spin-orbit coupling. In general case, the eigenenergies can be found only numerically. Here, we consider the limit $\Delta_2\gg\Delta_1,\lambda$, which is realized for majority of transition metal compounds. In addition to that, we are interesting in the low-energy sector $|\bar{\mu},\sigma\rangle=\{|e_{g1}^{\pi},\sigma\rangle,\,|e_{g2}^{\pi},
\sigma\rangle,\,|a_{1g},\sigma\rangle\}$. The projection of $\hat{H}$ to this reduced basis decreases the rank of the matrix to $6$. As a result, we are able to obtain analytical expressions for the eigenenergies and eigenfunctions. There are three doublets with energies (see Fig.~\ref{FigSO})
\begin{eqnarray}
E_1&=&2\left(\Delta_1-\Delta\right)\,,\nonumber\\
E_2&=&\Delta-\sqrt{\Delta^2+\xi^2}\,,\label{levels_SO}\\
E_3&=&\Delta+\sqrt{\Delta^2+\xi^2}\,,\nonumber
\end{eqnarray}
where
\begin{equation}
\Delta=\frac{\Delta_1}{2}+\frac{1+3\cos\alpha}{8}\lambda\,,
\;\;\xi=\sqrt{\frac{3}{2}}\sin\!\frac{\alpha}{2}\,\lambda\,.
\end{equation}
The eigenfunctions $|v_{1,2}^{(s)}\rangle$ corresponding to energies $E_s$ ($s=1,2,3$) are the following
\begin{eqnarray}
|v_{1}^{(1)}\rangle&=&|v_1,\uparrow\rangle\,,\nonumber\\
|v_{2}^{(1)}\rangle&=&|v_2,\downarrow\rangle\,,\label{v1}
\end{eqnarray}
\begin{eqnarray}
|v_{1}^{(2)}\rangle&=&\cos\delta|a_{1g},\uparrow\rangle+
\sin\delta|v_1,\downarrow\rangle\,,\nonumber\\
|v_{2}^{(2)}\rangle&=&\cos\delta|a_{1g},\downarrow\rangle+
\sin\delta|v_2,\uparrow\rangle\,,\label{v2}
\end{eqnarray}
\begin{eqnarray}
|v_{1}^{(3)}\rangle&=&-\sin\delta|a_{1g},\uparrow\rangle+
\cos\delta|v_1,\downarrow\rangle\,,\nonumber\\
|v_{2}^{(3)}\rangle&=&-\sin\delta|a_{1g},\downarrow\rangle+
\cos\delta|v_2,\uparrow\rangle\,,\label{v3}
\end{eqnarray}
where
\begin{equation}
|v_{1,2},\sigma\rangle=\frac{1}{\sqrt{2}}\left(i|e_{g1}^{\pi},
\sigma\rangle\mp|e_{g2}^{\pi},\sigma\rangle\right)\,,
\end{equation}
and
\begin{equation}\label{delta}
\tan2\delta=\frac{\xi}{\Delta}\,.
\end{equation}
Note, that while energies of the doublets, $E_s$, are the same for M$_1$ and M$_2$ magnetic ions, the corresponding eigenfunctions are different due to $\pm$ sign in Eq.~\eqref{t2g_c12_gen} defining the eigenfunctions of the $e_g^{\pi}$ levels. Note also, that Eqs.~\eqref{levels_SO}\,--\,~\eqref{delta} were obtained in the limit of $\Delta_2\to\infty$. The corrections to this result from the $e_g^{\sigma}$ sector can be found in perturbation theory on $E_s/\Delta_2$.

\begin{figure}[t]
\centering
\includegraphics[width=0.32\columnwidth]{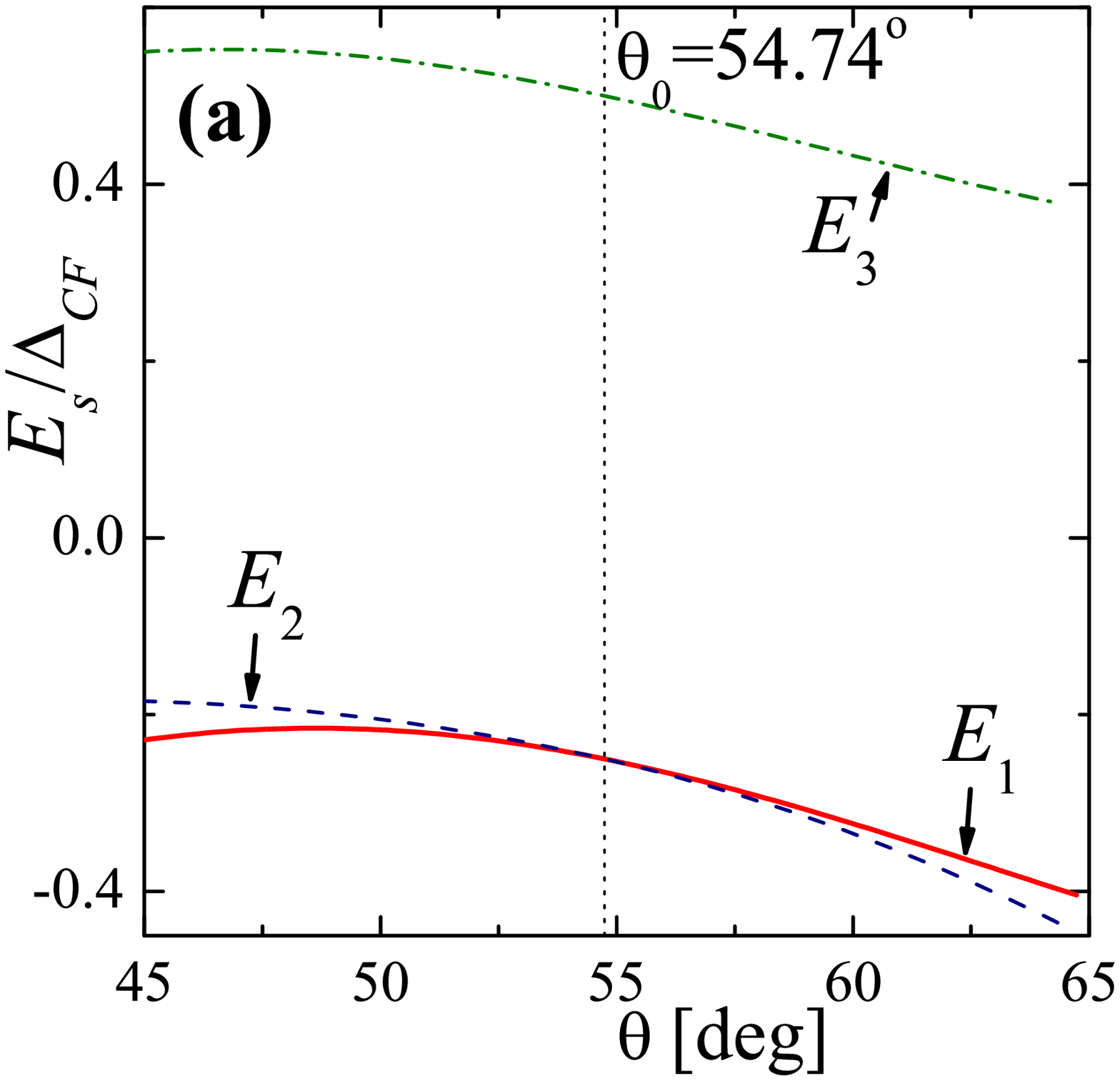}
\includegraphics[width=0.32\columnwidth]{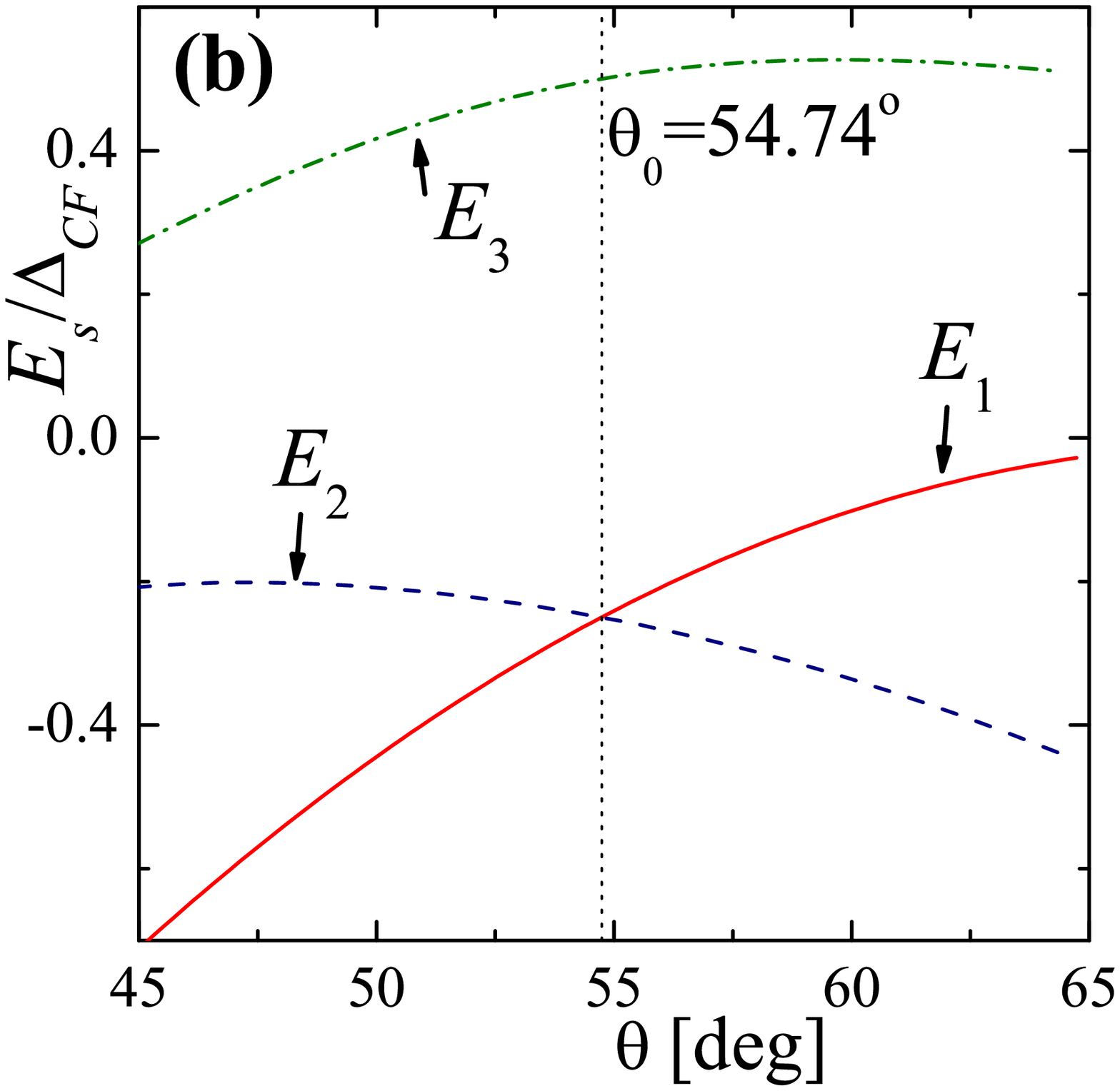}
\includegraphics[width=0.32\columnwidth]{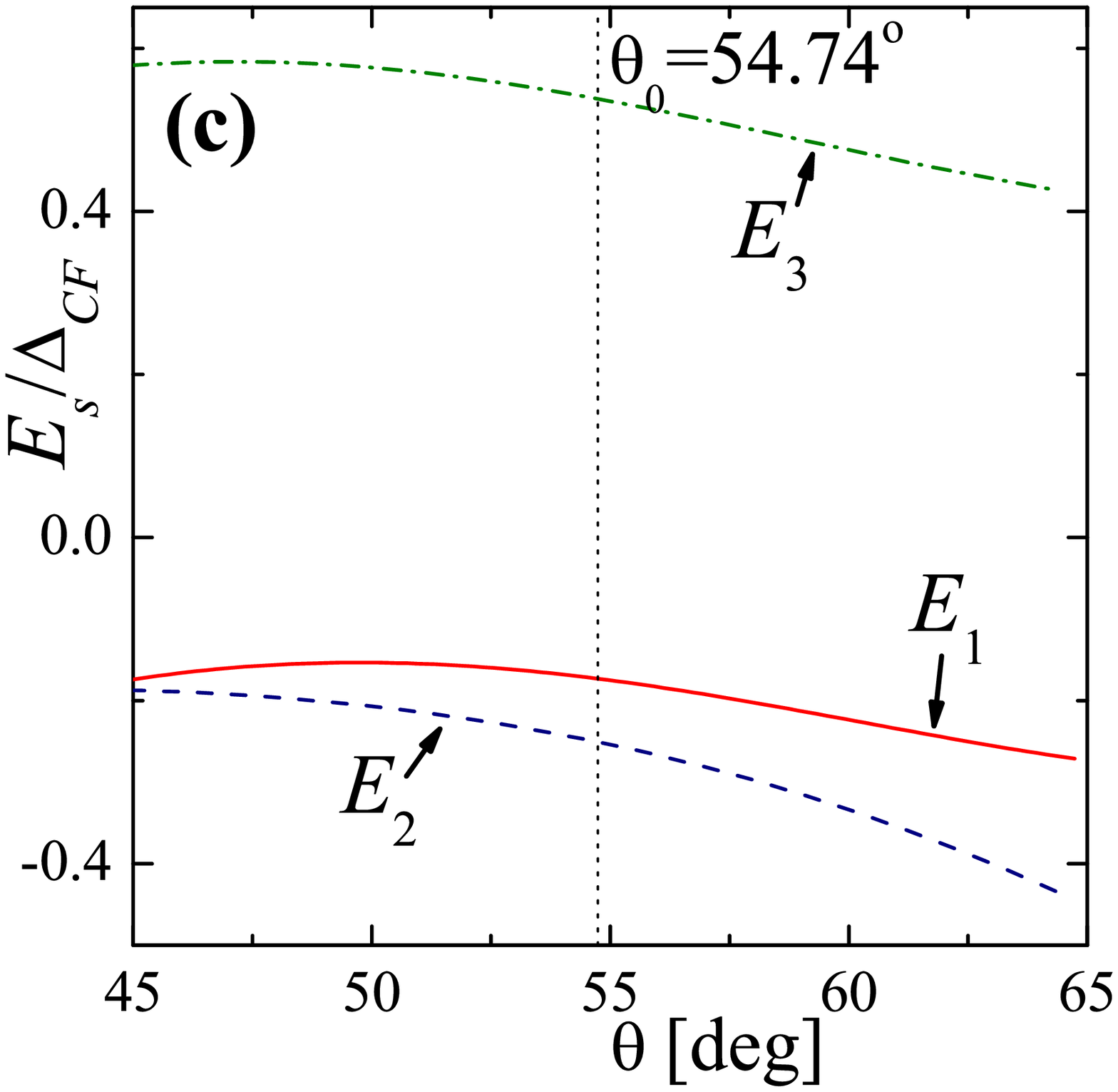}
\caption{(Color online) The energies of the Kramers doublets, $E_1$, $E_2$, and $E_3$, on the angle $\theta$ calculated for $\kappa=0.1$, $Z^{*}=0$ (panel {\it a}), $\kappa=1$, $Z^{*}=0$ (panel {\it b}), and $\kappa=0.1$, $Z^{*}=3$ (panel {\it c}). For all panels $\lambda/\Delta_{CF}=0.5$. For the definition of $\kappa$ and $Z^{*}$, see the text.}\label{FigELambda}
\end{figure}

Formulas~\eqref{levels_SO}\,--\,~\eqref{delta} are simplified in the limits of small and large spin--orbit coupling in comparison to the trigonal splitting. In the former case, when $\lambda\ll\Delta_1$, we obtain for energy levels up to the first order on $\lambda/\Delta_1$
\begin{equation}\label{levels_SO_small}
E_{1}\cong\Delta_1-\frac{1+3\cos\alpha}{4}\lambda\,,\;E_{2}\cong\Delta_1
+\frac{1+3\cos\alpha}{4}\lambda\,,\;E_3\cong0\,.
\end{equation}
In opposite limit, $\lambda\gg\Delta_1$, we obtain in the leading order
\begin{equation}\label{levels_SO_large}
E_1\!\cong\!-\frac{1+3\cos\alpha}{4}\lambda\,,\;
E_2\!\cong\!-\frac{3-3\cos\alpha}{4}\lambda\,,\;
E_3\!\cong\!\lambda\,.
\end{equation}
In this case, the parameter $\delta$ in Eqs.~\eqref{v2} and~\eqref{v3} is equal to
\begin{equation}\label{delta0}
\tan\delta=\sqrt{\frac{3}{2}}\sin\!\frac{\alpha}{2}\,.
\end{equation}
For the case of ideal octahedra and negligible effect of nearest-neighbor magnetic atoms [$\alpha=\alpha_0=\arccos(1/3)$, $\Delta_1\equiv0$], the formulas for $E_s$ and $\delta$ are simplified even further
\begin{equation}
E_1=E_2=-\frac{\lambda}{2}\,,\;E_3=\lambda\,,\;\;\delta=\arcsin(1/\sqrt{3})\,.
\end{equation}
Thus, in this case we have the low energy (if $\lambda>0$) quartet and higher energy doublet. In such a situation, we can introduce the effective angular momentum for $t_{2g}$ levels, $l_{eff}=1$, and the effective Hamiltonian becomes $H_{eff}=-\lambda\mathbf{l}_{eff}\mathbf{S}$. Note the opposite sign of spin-orbit coupling in comparison to the original Hamiltonian, Eq.~\eqref{H_SO}.

Let us consider now the behavior of the Kramers doublets for the trigonal distortions. The dependence of $E_1$, $E_2$, and $E_3$ on the angle $\theta$ calculated for $\lambda/\Delta_{CF}=0.5$ and different system's parameters $\kappa$ and $Z^{*}$ are shown in Fig.~\ref{FigELambda}. Analysis shows that for considerably large $\lambda$, energy $E_3$ lies above energies $E_1$ and $E_2$. If we neglect the effect of the neighboring magnetic ions to the crystal field ($Z^{*}=0$), we obtain that $E_1<E_2$ for the stretched ($\theta<\theta_0$) octahedra, and $E_1>E_2$ for the compressed ($\theta>\theta_0$) octahedra [see Fig.~\ref{FigELambda}(a,b)]. The contribution of the magnetic ions to the crystal field tends to make $E_1>E_2$ [see Fig.~\ref{FigELambda}(c)].

Considering each of three doublets separately, it is easy to demonstrate that the electron hopping integrals between the corresponding wave functions again meet the condition $t_{11} =t_{22} = t, t_{12}= 0$ and hence the exchange has the Heisenberg form, $H \sim J \sum{{\bm \sigma}_i{\bm \sigma}_j}$, where ${\bm \sigma}$ is the effective spin describing the $j=1/2$ doublet. However, ${\bm \sigma}$ has its own physical meaning for each doublet. This is always true for one electron or hole at $E_3$ level, but in the case of $E_1$ or $E_2$, the doublets should be far enough from the level-crossing point.

\end{document}